\documentclass{./jfm}

\usepackage[british]{babel}

\usepackage{graphicx}
\usepackage{epstopdf}

\usepackage{amsmath}
\usepackage{mathtools}
\usepackage[euler]{textgreek}

\usepackage{color}

\shorttitle{Bounds on heat transfer for infinite-{\it Pr} B\'enard--Marangoni convection}
\shortauthor{G. Fantuzzi, A. Pershin and A. Wynn}

\title{Bounds on heat transfer for B\'{e}nard--Marangoni convection at infinite Prandtl number}

\author{Giovanni Fantuzzi\aff{1}
  \corresp{\email{gf910@ic.ac.uk}},
  Anton Pershin\aff{2}
  \and
  Andrew Wynn\aff{1}}

\affiliation{
\aff{1}
Department of Aeronautics, Imperial College London, South Kensington Campus, London SW7 2AZ, U.K.
\aff{2}
Department of Applied Mathematics, University of Leeds, Leeds LS2 9JT, U.K. 
}

\newcommand{\BM}{B\'enard--Marangoni}
\newcommand{\RB}{Rayleigh--B\'enard}
\newcommand{\HD}{Hagstrom \& Doering}
\newcommand\Ma{\mbox{\textit{Ma}}}  
\newcommand\Nu{\mbox{\textit{Nu}}}  
\newcommand{\dx}{\,\mathrm{d}x}
\newcommand{\dz}{\,\mathrm{d}z}
\newcommand{\dt}{\,\mathrm{d}t}
\newcommand{\dxi}{\,\mathrm{d}\xi}
\newcommand{\eps}{\varepsilon}
\newcommand{\norm}[1]{\left\| #1 \right\|}
\newcommand{\normtwo}[1]{\left\| #1 \right\|_2}
\newcommand{\abs}[1]{\left\vert #1 \right\vert}
\newcommand{\ie}{{\it i.e.}}
\renewcommand{\vec}[1]{\boldsymbol{#1}}
\newcommand{\mat}[1]{\mathsfbi{#1}}
\newcommand*{\defeq}{\mathrel{\vcenter{\baselineskip0.5ex \lineskiplimit0pt
                     \hbox{.}\hbox{.}}}%
                     =}
\usepackage{hyperref}
\hypersetup{colorlinks=true,linkcolor=blue,urlcolor=blue,citecolor=blue}

\usepackage{natbib}
\makeatletter

\patchcmd{\NAT@citex}
  {\@citea\NAT@hyper@{%
     \NAT@nmfmt{\NAT@nm}%
     \hyper@natlinkbreak{\NAT@aysep\NAT@spacechar}{\@citeb\@extra@b@citeb}%
     \NAT@date}}
  {\@citea\NAT@nmfmt{\NAT@nm}%
   \NAT@aysep\NAT@spacechar\NAT@hyper@{\NAT@date}}{}{}

\patchcmd{\NAT@citex}
  {\@citea\NAT@hyper@{%
     \NAT@nmfmt{\NAT@nm}%
     \hyper@natlinkbreak{\NAT@spacechar\NAT@@open\if*#1*\else#1\NAT@spacechar\fi}%
       {\@citeb\@extra@b@citeb}%
     \NAT@date}}
  {\@citea\NAT@nmfmt{\NAT@nm}%
   \NAT@spacechar\NAT@@open\if*#1*\else#1\NAT@spacechar\fi\NAT@hyper@{\NAT@date}}
  {}{}

\makeatother

\begin{document}%
\maketitle

\begin{abstract}
The vertical heat transfer in {\BM} convection of a fluid layer with infinite Prandtl number is studied by means of upper bounds on the Nusselt number {\Nu} as a function of the Marangoni number {\Ma}. Using the background method for the temperature field, it has recently been proven by {\HD} that $\Nu\leq 0.838\,\Ma^{2/7}$. In this work we extend previous background method analysis to include balance parameters and derive a variational principle for the bound on {\Nu}, expressed in terms of a scaled background field, that yields a better bound than \HD's formulation at a given {\Ma}. Using a piecewise-linear, monotonically decreasing profile we then show that $\Nu\leq 0.803\,\Ma^{2/7}$, lowering the previous prefactor by 4.2\%. However, we also demonstrate that optimisation of the balance parameters does not affect the asymptotic scaling of the optimal bound achievable with \HD's original formulation. We subsequently utilise convex optimisation to optimise the bound on {\Nu} over all admissible background fields, as well as over two smaller families of profiles constrained by monotonicity and convexity. The results show that $\Nu\leq O(\Ma^{2/7}(\ln\Ma)^{-1/2})$ when the background field has a non-monotonic boundary layer near the surface, while a power-law bound with exponent 2/7 is optimal within the class of monotonic background fields. Further analysis of our upper-bounding principle reveals the role of non-monotonicity, and how it may be exploited in a rigorous mathematical argument.
\end{abstract}


\vspace*{-4em}
\section{Introduction}
\label{s:introduction}


When the surface of a layer of fluid experiences sufficiently strong local variations in temperature, surface-tension-induced shear stresses drive  bulk convective motion. {\BM} convection, as it is commonly known, arises in a variety of industrial processes, including drying of thin polymer films~\citep{Yiantsios2015}, fusion welding~\citep{Debroy1995}, laser cladding~\citep{Kumar2009}, and the growth of single-crystal semiconductors~\citep[Chapter 3 and references therein]{Lappa2010}. Shear-driven convection is also observed in distillation columns~\citep{Zuiderweg1958,Patberg1983} and in differentially heated fluids in microgravity environments, where buoyancy effects are negligible~\cite[Chapter 2]{Lappa2010}. 

Despite its widespread applications, the dynamics and heat transfer properties of {\BM} convection have been studied far less than those of buoyancy-driven {\RB} convection. One fundamental question that remains largely unanswered is how the net vertical heat transfer across the layer, described by the Nusselt number {\Nu}, depends on the external forcing, measured by the Marangoni number {\Ma}. A phenomenological bounday layer scaling analysis put forward by~\cite{Pumir1996} predicts a transition from $\Nu=O(\Ma^{1/4})$ to $\Nu=O(\Ma^{1/3})$ as laminar convection rolls are replaced by turbulent convection, with prefactors that depend on the Prandtl number {\Pran} --- the ratio of the fluid's kinematic viscosity and its thermal diffusivity. Two-dimensional direct numerical simulations (DNSs) at low {\Pran} and large {\Ma}~\citep{Boeck1998,Boeck2005} confirm the $1/3$ scaling exponent for the turbulent regime when free-slip conditions are imposed on the velocity field, but $\Nu = O(\Ma^{1/5})$ is observed in the no-slip case. Moreover, further DNSs by~\citet{Boeck2001} indicate that {\BM} convection in high-Prandtl-number fluids may not be turbulent even when {\Ma} is $10^4$ times the value at which convection first appears. Under the assumption that the observed stationary convection rolls remain stable as {\Ma} is raised when {\Pran} is infinite, the same authors predict that $\Nu = O(\Ma^{2/9})$ in this limit.

Unfortunately, available experimental data~\citep[see][and references therein]{Schatz2001,Eckert2006} do not reach the highly nonlinear regime, where these scaling laws are thought to apply. An alternative approach to confirm or disprove them is to try and derive rigorous bounds on {\Nu} as a function of {\Ma} directly from the governing equations. This can be done without recourse to statistical hypothesis or closure models using the background method~\citep{Doering1992,Doering1994,Doering1996,Constantin1995,Constantin1995a}. The essence of the method is to write the temperature field as the sum of a steady ``background'' component $\tau$ and a time-dependent fluctuation, and show that if $\tau$ satisfies a particular nonlinear stability condition, then {\Nu} is bounded as a function of $\tau$ only. The problem that results is variational in nature: optimise the bound on {\Nu} over all stable background fields. 

The background method has been applied extensively to the {\RB} problem in a variety of configurations~\citep[see e.g.][]{Doering1996,Otero2002,Doering2006,
Wittenberg2010a,Whitehead2011,Whitehead2012,Goluskin2016d}.
On the other hand, the only result for {\BM} convection is due to \citet{Hagstrom2010}, who used a monotonically decreasing, piecewise-linear  background temperature field to prove 
$\Nu\leq 0.841\times\Ma^{1/2}$ 
for finite-Prandtl-number fluids, while $\Nu\leq 0.838\times\Ma^{2/7}$ in the infinite-{\Pran} limit. 
%

This work investigates whether \HD's bound for {\BM} convection at infinite Prandtl number can be lowered, reducing the gap with the DNS results and phenomenological predictions of~\citet{Boeck2001}. The assumption of infinite {\Pran} significantly simplifies the mathematical treatment of the problem, making it amenable to analysis, and still provides an accurate model for large-{\Pran} fluids~\citep{Boeck2001}, including some silicone oils used in experiments~\citep{DeBruyn1996}. 
%

Our primary aim is to determine the best possible upper bound on {\Nu} when the background method is applied to the temperature field. To this end, we revisit \HD's background method analysis and derive a new upper-bounding variational principle for the Nusselt number that includes two so-called ``balance parameters''~\citep{Nicodemus1998c}. One of these balance parameters can be optimised analytically, while the remaining one and the background temperature field can be combined to formulate a bound on {\Nu} in terms of a scaled background profile. We then employ convex programming to optimize the scaled background field for Marangoni numbers up to $\Ma = 10^9$, and observe that the optimal bounds take the form  $\Nu\leq O(\Ma^{2/7}(\ln\Ma)^{-1/2})$---a logarithmic improvement on \HD's bound.

We also seek to identify which features of the optimal scaled background temperature field are key to lowering the bound on {\Nu}. For instance, non-monotonicity plays an important role in the background method analysis for infinite-{\Pran} {\RB} convection~\citep{Plasting2005,Doering2006}, and it is natural to ask if the same is true for the {\BM} problem. Another important issue is whether one can expect to improve \HD's bound using a relatively simple background field, which is amenable to rigorous mathematical analysis. To answer these questions we utilise convex optimisation once again and minimise the bound on {\Nu} over two families of scaled background fields: those that decrease monotonically, and those constrained by convexity. Our results are supported by analysis of the variational principle for the bound, which also suggests a way to proceed with a rigorous mathematical proof.



Numerical optimisation of the bound on {\Nu} is central to this work, and our computational strategy deserves some remarks.
Traditionally, the Euler--Lagrange equations for the optimal background field and balance parameters are derived, discretised, and solved~\citep[see e.g.][]{Plasting2003,Wen2013,Wen2015a}. Instead, we discretise the variational problem for the bound to obtain a convex conic programme, {\ie}, a convex optimisation problem in which the variables are constrained to belong to a convex cone. 
The procedure is similar to that described in previous works by the authors~\citep{Fantuzzi2015,Fantuzzi2016PRE}, however here we use a different discretisation method. The first advantage of this approach is that very efficient software packages are available to solve conic programmes. The second is that additional linear constraints on the background field, such as monotonicity and convexity, can be included in a straightforward way and without any changes to the numerical optimisation algorithm. Conic programming, therefore, enables one to interrogate the bounding principle in a systematic way, in order to inform rigorous mathematical analysis. This applies  not only to infinite-{\Pran} {\BM} convection, but to any convex upper-bounding variational problem obtained from the application of the background method. 



The outline of this work is the following. Section~\ref{s:model} introduces Pearson's model~\citep{Pearson1958} for {\BM} convection at infinite Prandtl number, which is our starting point. We apply the background method with balance parameters to formulate an uper-bounding variational principle for the Nusselt number in \S\ref{s:BackgroundMethod}, and compare it to the one derived by~\cite{Hagstrom2010} in \S\ref{s:HagstromDoering}. Section~\ref{s:numerics} is devoted to the numerical optimisation of the background fields, and describes our computational approach in detail. We discuss our results in \S\ref{s:discussion} with the help of additional analysis of the variational problem for the bound. Section~\ref{s:conclusion} concludes the paper.

Our notation will be mostly standard. Upon non-dimensionalising, we consider a two-dimensional, horizontally-periodic layer with domain $[0,2\pi]\times[0,1]$, with $x$ and $z$ denoting the horizontal and vertical coordinates, respectively. The $L^2$ and $L^\infty$ norms in the $z$ direction will be denoted by $\normtwo{\cdot}$ and $\norm{\cdot}_\infty$, respectively, {\ie}
\begin{align}
\normtwo{q} &\defeq \left( \int_0^1 \vert q(z,\cdot) \vert^2 \dz \right)^{1/2}, &
\norm{q}_\infty &\defeq \sup_{z\in[0,1]} \vert q(z,\cdot) \vert.
\end{align}
Overlines denote horizontal and infinite-time averages, while angle brackets indicate volume and infinite-time averages, {\ie}
%
\begin{align}
\overline{q}(z) &\defeq
\lim_{\mathcal{T}\to\infty} \frac{1}{\mathcal{T}} \int_0^\mathcal{T}\!
\frac{1}{2\pi} \int_0^{2\pi} q(x,z,t) \dx\,\mathrm{d}t,
&
\left\langle q \right\rangle &\defeq
\int_0^1\overline{q}(z)\dz.
\end{align}
Since infinite-time averages need not exist in general, one could be more rigorous and replace $\lim$ with $\limsup$. Note also that $\langle \abs{q(z)}^2 \rangle = \normtwo{q}^2$ when $q$  depends only on $z$. 
%

\section{Pearson's model}
\label{s:model}

Consider a two-dimensional layer of incompressible fluid of depth $h$, density $\rho$, kinematic viscosity $\nu$, thermal diffusivity $\kappa$ and thermal conductivity $\lambda$~\citep[the model and the results may be generalised to the three-dimensional case as described in][]{Hagstrom2010}. The fluid is heated from below at constant temperature, and cooled at the surface with a fixed heat flux $q$. The problem is made non-dimensional using $h$ as the length unit, $h^2/\kappa$ as the time unit, and $qh/\lambda$ as the temperature unit. When the Prandtl number $\Pran = \nu/\kappa$ is infinite, Pearson's equations for the fluid's motion~\citep{Pearson1958} reduce to~\citep{Hagstrom2010} 
\begin{subequations}
\begin{align}
\label{e:NSmomentum}
\bnabla p &= \bnabla^2\vec{u},
\\
\label{e:NStemp}
\partial_t T
+ \vec{u}\bcdot\bnabla T &= \bnabla^2 T,
\\
\label{e:continuity}
\bnabla \bcdot \vec{u} &= 0,
\end{align}
\end{subequations}
where $\vec{u}(x,z,t)=u(x,z,t)\vec{i}+w(x,z,t)\vec{k}$ is the fluid's velocity, $p(x,z,t)$ is the pressure, and $T(x,z,t)$ is the temperature. 
All variables are assumed to be periodic in the horizontal direction ({\ie} along the $x$ axis) with period $2\pi$, and satisfy the vertical boundary conditions (BCs)
\begin{align}
\label{e:BC}
\vec{u}\vert_{z=0} &= 0, &
w\vert_{z=1}&=0,  &
T\vert_{z=0}&=0,  &
\partial_z T\vert_{z=1} &= -1.
\end{align}
The fluid is driven at the top boundary by surface tension forces due to local temperature gradients, which induce motion in the bulk of the layer through the action of viscosity. Mathematically, the situation is described by the additional BC
\begin{equation}
\label{e:BCshear}
\left[ \partial_z u + \Ma\,\partial_x T\right]_{z=1}=0.
\end{equation}
The Marangoni number 
$\Ma=\gamma q h^2/(\lambda \rho \nu \kappa)$, where $\gamma$ is the negative of the derivative of the surface tension with respect to the fluid's temperature,
describes the ratio of surface tension to viscous forces, and is the governing non-dimensional parameter of the flow.

\begin{figure}
\centerline{
\includegraphics{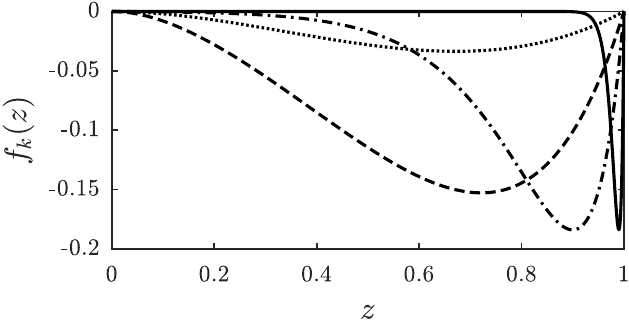}
}
\caption{The function $f_k(z)$ for $k=1$ (dotted line), $k=3$ (dashed line), $k=10$ (dot-dashed line), and $k=100$ (solid line).}
\label{f:fkfigure}
\end{figure}

The purely conductive state $\vec{u}(x,z,t)=0$, $p=\text{constant}$, $T(x,z,t) = -z$
is asymptotically stable when $\Ma\leq 66.84$~\citep{Fantuzzi2017a}, while for $\Ma\geq 79.61$ it is subject to linear instabilities~\citep{Pearson1958} and convection sets in~\citep{Boeck1998,Boeck2001}. Taking the divergence of~\eqref{e:NSmomentum} and using incompressibility shows that $\nabla^2p=0$, so taking the Laplacian of~\eqref{e:NSmomentum} gives
\begin{equation}
\label{e:BiHarmonicU}
\nabla^4 \vec{u} = 0.
\end{equation}
Thus, each component of the ensuing convective velocity is bi-harmonic, and can be determined as a linear function of the temperature field, which forces~\eqref{e:BiHarmonicU} via the BC~\eqref{e:BCshear}.  In particular the horizontal Fourier coefficients $\hat{w}_k(z)$, $k\in\mathbb{Z}$, of the vertical velocity $w$ can be computed as a function of the horizontal Fourier coefficients $\hat{T}_k(z)$ of the temperature. One finds~\citep{Hagstrom2010}
\begin{equation}
\label{e:wDef}
\hat{w}_k(z) = -\Ma\,f_k(z)\,\hat{T}(1), \qquad k\in\mathbb{Z},
\end{equation}
where $f_0(z)=0$ (so $\hat{w}_0=0$ and $w$ has zero horizontal mean), and
\begin{equation}
\label{e:fk}
f_k(z) = \frac{k\sinh k \left[ kz\cosh(kz) - \sinh(kz) + (1-k\coth k)\,z \sinh(kz)\right]}{\sinh(2k)-2k} 
, 
\quad k\in\mathbb{Z}\setminus\{0\}.
\end{equation}
Note that the function $f_k$ satisfies $f_k(z)\leq 0$ for $z\in[0,1]$, $f_k(0)=0=f_k(1)$, and $f_k(z)\to 0$ pointwise for all $z\in(0,1)$ as $k\to\infty$ (see figure~\ref{f:fkfigure}; note that the corresponding figure in Hagstrom \& Doering's original paper is incorrect: they plot the {\it negative} of $f_k$). 

%

Convection enhances the vertical heat transport, and since the BC $\partial_z T\vert_{z=1}=-1$ prescribes the heat flux through the top surface, the net effect is a reduction in the temperature drop across the layer. The key non-dimensional parameter to quantify this process is the Nusselt number
\begin{equation}
\label{e:NuDef}
\Nu \defeq-\frac{1}{\overline{T}(1)} 
= \frac{1}{\langle \abs{\bnabla T}^2 \rangle},
\end{equation} 
where $\abs{\bnabla T}^2=(\partial_x T)^2 + (\partial_z T)^2$. The first equality in~\eqref{e:NuDef} defines the Nusselt number, while the second one can be proven by taking the volume and infinite-time average of $T\times$\eqref{e:NStemp}, followed by appropriate integrations by parts using~\eqref{e:continuity} and the BCs~\citep[for more details, see][]{Hagstrom2010}.

\section{An upper-bounding variational principle for the Nusselt number}
\label{s:BackgroundMethod}


\subsection{The background method with balance parameters}
\label{s:BM_BalanceParameters}
The background method analysis begins by decomposing the temperature variable as 
\begin{equation}
\label{e:BFdecomposition}
T(x,z,t)=\tau(z) + \theta(x,z,t),
\end{equation}
where the steady background field $\tau(z)$ satisfies the BCs
\begin{align}
\label{e:bfBC}
\tau(0)=0, \qquad \tau'(1) = -1,
\end{align}
while the time-dependent perturbation $\theta(x,z,t)$ is periodic in the horizontal direction and satisfies 
\begin{equation}
\label{e:pertBC}
\theta\vert_{z=0} = 0, \qquad
\partial_z\theta \vert_{z=1} = 0.
\end{equation}

Upon substituting this decomposition into~\eqref{e:NSmomentum} we obtain an evolution equation for the perturbation $\theta$,
\begin{equation}
\label{e:PertEq}
\partial_t \theta + \vec{u}\bcdot\bnabla \theta = \bnabla^2 \theta + \tau'' - w\,\tau'.
\end{equation}
Averaging $\theta\times$\eqref{e:PertEq} over the volume and infinite time, followed by appropriate integration by parts using~\eqref{e:continuity} and the BCs for $\theta$ in~\eqref{e:pertBC}, shows that 
\begin{equation}
\label{e:PertEnEq}
\left\langle 
\vert \bnabla \theta \vert^2 + \tau'\,\partial_z\theta + \tau'\,w\,\theta
\right\rangle + \overline{\theta}(1) =0.
\end{equation}
Moreover, substituting~\eqref{e:BFdecomposition} into~\eqref{e:NuDef} gives the two identities
\begin{subequations}
\begin{align}
\label{e:NuId1}
\Nu^{-1} + \overline{\theta}(1) + \tau(1) &= 0,
\\
\label{e:NuId2}
\Nu^{-1} - \left\langle \vert \bnabla \theta \vert^2 + 2\,\tau'\,\partial_z\theta\right\rangle - \normtwo{\tau'}^2 &= 0.
\end{align}
\end{subequations}
%
Taking the linear combination $\alpha\times$\eqref{e:PertEnEq}$-\beta\times$\eqref{e:NuId1}$+$\eqref{e:NuId2} for scalar balance parameters $\alpha,\beta\neq 1$ to be determined, using the fact that
$\overline{\theta}(1) = \langle \partial_z \theta \rangle$
by virtue of~\eqref{e:pertBC}, and rearranging yields 
\begin{equation}
\label{e:lincomb}
\frac{1}{\Nu} = 
-\frac{\normtwo{\tau'}^2+\beta\,\tau(1)}{\beta-1}
+ \frac{\alpha-1}{\beta-1}\,
\mathcal{Q}\{\theta,w\},
\end{equation}
where
\begin{equation}
\label{e:SCOriginal}
\mathcal{Q}\{\theta,w\}= 
	\left\langle
	\vert \bnabla \theta \vert^2 
    	+\frac{\alpha}{\alpha-1}\,\tau'\,w\,\theta 
    +\left(  \frac{\alpha-2}{\alpha-1}\,\tau' 
    			+\frac{\alpha-\beta}{\alpha-1}
    	\right) \partial_z \theta
    \right\rangle.
\end{equation}
If the balance parameters are chosen to satisfy
\begin{equation}
\label{e:PosAlphaBeta}
\frac{\alpha-1}{\beta-1} > 0
\end{equation}
we can bound
\begin{equation}
\label{e:NuBound1}
\frac{1}{\Nu} \geq 
-\frac{\normtwo{\tau'}^2+\beta\,\tau(1)}{\beta-1}
+ \frac{\alpha-1}{\beta-1}
\,\inf_{\theta,\,w}\mathcal{Q}\{\theta,w\},
\end{equation}
where the infimum is taken over all horizontally periodic fields $\theta$ that satisfy the BCs in~\eqref{e:pertBC} and over all velocity fields $w$ with horizontal Fourier coefficients given by~\eqref{e:wDef}. The key simplification is that we do not require $\theta$ to satisfy the nonlinear evolution equation~\eqref{e:PertEq}. As a result, we may without any loss of generality restrict our attention to time-independent perturbations, and interpret $\langle \cdot \rangle$ in~\eqref{e:SCOriginal} as a volume average. 

To compute the infimum in~\eqref{e:NuBound1} we substitute the Fourier expansions for $\theta$ and $w$ into~\eqref{e:SCOriginal}. Noticing that $\hat{\theta}_k=\hat{T}_k$ for $k\neq 0$ by virtue of~\eqref{e:BFdecomposition}, and that $f_0(\cdot)=0$ in~\eqref{e:wDef}, the Fourier coefficients $\hat{w}_k$ can be expressed in terms of $\hat{\theta}_k$ as
\begin{equation}
\label{e:wDef_theta}
\hat{w}_k(z) = -\Ma\,f_k(z)\,\theta_k(1), \qquad k\in\mathbb{Z}.
\end{equation} 
Moreover, $\hat{\theta}_{-k} = \hat{\theta}_k^*$  (where $^*$ denotes complex conjugation) because the Fourier modes must combine into the real-valued temperature perturbation $\theta$. Consequently, we may rewrite
\begin{equation}
\mathcal{Q}\{\theta,w\} = \mathcal{Q}_0\{\hat{\theta}_0\} + 2\,\sum_{k\geq 1} \mathcal{Q}_k\{\hat{\theta}_k\}
\end{equation}
where
\begin{equation}
\mathcal{Q}_0\{\hat{\theta}_0\} \defeq \int_0^1 
\left[\abs{{\hat{\theta}_0}'(z)}^2 + \left(  \frac{\alpha-2}{\alpha-1}\,\tau'(z) 
+\frac{\alpha-\beta}{\alpha-1}
\right) {\hat{\theta}_0}'(z) \right] \dz,
\end{equation}
while for $k\geq 1$ the last term in~\eqref{e:SCOriginal} vanishes and we have
\begin{equation}
\label{e:QkDef}
\mathcal{Q}_k\{\hat{\theta}_k\} \defeq \int_0^1 \left\{ \abs{{\hat{\theta}_k}'(z)}^2 +k^2\abs{\hat{\theta}_k(z)}^2 
-\frac{\alpha\,\Ma}{\alpha-1}\,\tau'(z)\,f_k(z)\,\Real\!\left[\hat{\theta}_k(1)\,{\hat{\theta}_k(z)}^*\right]
\right\} \dz.
\end{equation}
%
%
Now, the infimum of $\mathcal{Q}\{\theta,w\}$ must be negative semidefinite since $\mathcal{Q}\{0,0\}=0$. Moreover, each functional $\mathcal{Q}_k$, $k\geq 0$ must be individually lower bounded because among all perturbations $\theta$, $w$ are those with only one horizontal wavenumber. In light of~\eqref{e:pertBC}, this lower bound must be sought over all complex-valued functions $\hat{\theta}_k(z)$ that satisfy $\hat{\theta}_k(0)=0=\hat{\theta}_k'(1)$. Since $\mathcal{Q}_0\{0\}=0$, the infimum of $\mathcal{Q}_0$ must be negative semidefinite. When $k\geq 1$, instead,  $\mathcal{Q}_k$ is a homogeneous functional and so if it is lower bounded, its infimum must be exactly zero. Consequently,
\begin{equation}
\inf_{\theta,\,w}\mathcal{Q}\{\theta,w\} = 
\begin{cases}
\displaystyle\inf_{\hat{\theta}_0} \mathcal{Q}_0\{\hat{\theta}_0\} 
&\text{if } \mathcal{Q}_k\{\hat{\theta}_k\}\geq 0,\; k=1,\,2,\,\ldots,
\\
-\infty & \text{otherwise.}
\end{cases}
\end{equation}

In appendix~\ref{a:infQ0} we show that
\begin{equation}
\label{e:infQ0}
\inf_{\hat{\theta}_0} \mathcal{Q}_0\{\hat{\theta}_0\} =
-\frac{\normtwo{(\alpha-2)\tau' + \alpha-\beta}^2}{4(\alpha-1)^2}.
\end{equation}
Substituting this into~\eqref{e:NuBound1}, and using the fact that
\begin{equation}
\label{e:bfInt}
\tau(1) = \int_0^1 \tau'(z) \dz
\end{equation}
by virtue of~\eqref{e:bfBC} to simplify the resulting expression, we obtain
\begin{equation}
\label{e:NuBound2}
\frac{1}{\Nu} \geq
-\frac{4\,\alpha(\beta-1)\tau(1) + \normtwo{\alpha\tau'+\alpha-\beta}^2}{4(\alpha-1)(\beta-1)}.
\end{equation}

This bound is valid if~\eqref{e:PosAlphaBeta} holds, and if the background field $\tau$ is chosen to make the functional $\mathcal{Q}_k\{\hat{\theta}_k\}$ in~\eqref{e:QkDef} positive semidefinite for all (integer) wavenumbers $k\geq 1$.
%
The latter set of constraints can be combined into the single condition that
\begin{equation}
\label{e:SC}
\left\langle \abs{\nabla\theta}^2 + \frac{\alpha}{\alpha-1}\tau'\,w\,\theta\right\rangle \geq 0
\end{equation}
for all perturbations $\theta$, $w$ with {\it zero horizontal mean} that satisfy~\eqref{e:pertBC} and~\eqref{e:wDef_theta}. Using well-established terminology, we refer to such $\theta$ and $w$ as {\it admissible perturbations}, and to~\eqref{e:SC} as the {\it spectral constraint}.

The best possible bound on {\Nu} is then found upon solving the following optimization problem:
\begin{equation}
\label{e:varProbFull0}
\begin{aligned}
\sup_{\tau(z),\,\alpha,\,\beta} 
\quad &-\frac{4\,\alpha(\beta-1)\tau(1) + \normtwo{\alpha\tau'+\alpha-\beta}^2}{4(\alpha-1)(\beta-1)}
\\
\text{subject to} \quad 
&\left\langle \abs{\nabla\theta}^2 + \frac{\alpha}{\alpha-1}\tau'\,w\,\theta\right\rangle \geq 0 \quad \forall \text{ admissible } \theta,\,w,\\
&\frac{\alpha-1}{\beta-1}>0,\\
&\tau(0)=0,\\
&\tau'(1)=-1.
\end{aligned}
\end{equation}
Note that we look for the supremum of the objective function (rather than its maximum) because the strict inequality $(\alpha-1)/(\beta-1)>0$ may prevent the existence of a maximiser. 

\subsection{Optimization over \textbeta}
\label{s:optimizeBP}

The lower bound~\eqref{e:NuBound2} can be optimised over $\beta$ in a relatively straightforward way, because the spectral constraint is independent of $\beta$. Upon setting to zero the first derivative of the right-hand side of~\eqref{e:NuBound2} with respect to $\beta$, and using~\eqref{e:bfInt} to rearrange, we find two stationary values,
\begin{align}
\label{e:BetaPlusMinus}
\beta_{+} &= 1 + \normtwo{\alpha\,\tau'+\alpha-1}, &
\beta_{-} &= 1 - \normtwo{\alpha\,\tau'+\alpha-1}.
\end{align}
Inspection of the second derivative of the right-hand side of~\eqref{e:NuBound2} with respect to $\beta$ reveals that when $\alpha$ is constrained by~\eqref{e:PosAlphaBeta} both choices $\beta=\beta_+$ and $\beta=\beta_-$ correspond to a local maximum. Determining the optimal choice of $\beta$ therefore requires comparing the values of such local maxima. 

After choosing $\beta = \beta_+$ and re-parametrising 
$\alpha = \lambda/(\lambda-1)$---with $\lambda>1$  to satisfy~\eqref{e:PosAlphaBeta}---we can use~\eqref{e:bfInt} to rewrite~\eqref{e:NuBound2} as
%
\begin{equation}
\label{e:NuBoundOptBeta}
\frac{1}{\Nu} \geq \frac{1 - \normtwo{\lambda\,\tau'+1} - \lambda\,\tau(1)}{2}.
\end{equation}
The spectral constraint~\eqref{e:SC} can also be expressed in terms of $\lambda$ as
\begin{equation}
\label{e:SClambda}
\langle \abs{\nabla\theta}^2 + \lambda\,\tau'\,w\,\theta \rangle \geq 0 
\quad 
\forall \text{ admissible } \theta,\,w.
\end{equation}
Upon introducing the scaled background field $\rho(z) = \lambda\,\tau(z) = \alpha/(\alpha-1)\,\tau(z)$, subject to a suitably scaled version of the BCs in~\eqref{e:bfBC}, the optimal bound on {\Nu} corresponding to the choice $\beta=\beta_+$ is found by solving the variational problem
\begin{equation}
\label{e:varProbFullA}
\begin{aligned}
\sup_{\rho(z),\,\lambda} \quad & \frac{1 - \normtwo{\rho'+1} - \rho(1)}{2}\\
\text{subject to} \quad 
&\langle \abs{\nabla\theta}^2 + \rho'\,w\,\theta \rangle \geq 0 \quad \forall \text{ admissible } \theta,\,w,\\
&\rho(0)=0,\\
&\rho'(1)=-\lambda,\\
&\lambda>1.
\end{aligned}
\end{equation}

Similar steps show that the best possible bound on $\Nu$ when setting $\beta=\beta_{-}$ in~\eqref{e:NuBound2} is given by the solution of an optimisation problem that differs from~\eqref{e:varProbFullA} only in the constraint for $\lambda$,
\begin{equation}
\label{e:varProbFullB}
\begin{aligned}
\sup_{\rho(z),\,\lambda} \quad & \frac{1 - \normtwo{\rho'+1} - \rho(1)}{2},\\
\text{subject to} \quad 
&\langle \abs{\nabla\theta}^2 + \rho'\,w\,\theta \rangle \geq 0 \quad \forall \text{ admissible } \theta,\,w,\\
&\rho(0)=0,\\
&\rho'(1)=-\lambda,\\
&\lambda<1.
\end{aligned}
\end{equation}

The key observation at this stage is that the suprema in~\eqref{e:varProbFullA} and~\eqref{e:varProbFullB} coincide despite the different constraint on $\lambda$, and furthermore they are equal to the optimal value of the variational problem
\begin{equation}
\label{e:varProbFullC}
\begin{aligned}
\max_{\rho(z)} \quad & \frac{1 - \normtwo{\rho'+1} - \rho(1)}{2},\\
\text{subject to} \quad 
&\langle \abs{\nabla\theta}^2 + \rho'\,w\,\theta \rangle \geq 0 \quad \forall \text{ admissible } \theta,\,w,\\
&\rho(0)=0.
\end{aligned}
\end{equation}
In fact, for any value of $\lambda$ we can construct a feasible $\rho(z)$ for either~\eqref{e:varProbFullA} or~\eqref{e:varProbFullB} that approximates the solution of~\eqref{e:varProbFullC} arbitrarily accurately: simply let $\rho_0(z)$ be an $\varepsilon$-suboptimal strictly feasible point for~\eqref{e:varProbFullC}, and choose $\rho'(z)=\rho_0'(z)$ in~\eqref{e:varProbFullA} or~\eqref{e:varProbFullB} except for an infinitesimally thin layer near $z=1$, where $\rho'(z)=-\lambda$. A rigorous argument follows steps similar to those used in the energy stability analysis of the conductive state~\citep{Fantuzzi2017a}, and is omitted for brevity. The conclusion is satisfactory: the bound on $\Nu$ is independent of whether one sets $\beta=\beta_+$ or $\beta=\beta_-$ in~\eqref{e:NuBound2}.

%
\subsection{An explicit value for the optimal \textbeta}
\label{s:optimalBeta}

The variational principle~\eqref{e:varProbFullC} has been obtained by optimising the balance parameter $\beta$ as a function of the other balance parameter, $\alpha$, and the background field $\tau(z)$. Interestingly, the optimality conditions for the solution $\rho_\star(z)$ of~\eqref{e:varProbFullC} allow deriving a precise numerical value for the optimal $\beta$ even though the optimal $\alpha$ and $\tau(z)$ are unknown.
To show this, we introduce a variable $s$ such that $\normtwo{\rho'+1}\leq s$ and note that~\eqref{e:varProbFullC} is equivalent to
\begin{equation}
\label{e:varProbFullD}
\begin{aligned}
\max_{\rho(z), s} \quad & 1 - s - \rho(1),\\
\text{subject to} \quad 
&\langle \abs{\nabla\theta}^2 + \rho'\,w\,\theta \rangle \geq 0 \quad \forall \text{ admissible } \theta,\,w,\\
&\rho(0)=0,\\
&\normtwo{\rho'+1}\leq s.
\end{aligned}
\end{equation}
%
The feasible set of this problem is convex, so the linear objective function is maximised on the constraint boundary. Since for any given $\rho(z)$ we can always choose $s = \normtwo{\rho'+1}$, the optimal bound is attained when $\rho(z)$ is on the boundary of the feasible set of the spectral constraint, {\ie} when
\begin{equation}
\inf_{\theta,w\neq 0} \langle \abs{\nabla\theta}^2 + \rho'\,w\,\theta \rangle = 0.
\end{equation}
Since the spectral constraint is homogeneous in $\theta$ and $w$, it suffices to restrict our attention to admissible $\theta$ and $w$ satisfying some normalisation condition $\mathcal{N}\{\theta,w\}=0$ that excludes the zero fields. The optimal scaled background field $\rho_\star(z)$ and the optimal value $s_\star$ are then those that maximise the Lagrangian functional
\begin{multline}
\label{e:AugLag}
\mathcal{L}\{\rho,s,\theta,w,\zeta,\eta,\mu\} \defeq 1- s - \rho(1) 
+ \zeta\,\langle \abs{\nabla\theta}^2 + \rho'\,w\,\theta \rangle 
\\
+ \eta\,\left(s^2 - \normtwo{\rho'+1}^2\right)
+ \mu\,\mathcal{N}\{\theta,w\},
\end{multline}
where $\zeta$, $\eta$ and $\mu$ are scalar Lagrange multipliers. 

Setting to zero the first variation of $\mathcal{L}$ with respect to $\rho(z)$ 
shows that the optimal scaled background field $\rho_\star(z)$ must satisfy the ``natural'' boundary condition
\begin{equation}
\label{e:eta1}
1 + 2\,\eta + 2\,\eta\,\rho_\star'(1) = 0.
\end{equation}
(Of course, $\rho_\star(z)$ must also satisfy an Euler--Lagrange differential equation, but this will not be important here.) Moreover, setting to zero the derivatives of $\mathcal{L}$ with respect to $s$ and $\eta$, and eliminating $s$ yields
\begin{equation}
\label{e:eta2}
2\,\eta \normtwo{\rho_\star'+1} - 1 = 0.
\end{equation}
At this point, note that if $\rho'(z)=-1$ the spectral constraint~\eqref{e:varProbFullD} reduces to the condition for global ``energy'' stability of the conduction solution~\citep[see e.g.][]{Fantuzzi2017a}, which cannot be satisfied in the convective regime. Consequently, $\normtwo{\rho_\star'+1}\neq 0$ and we may use~\eqref{e:eta2} to eliminate $\eta$ from~\eqref{e:eta1}. The optimal scaled background field must therefore satisfy
\begin{equation}
\label{e:OptRhoCond1}
1 + \normtwo{\rho_\star'+1} + \rho_\star'(1) = 0.
\end{equation}

In particular, this implies that $-\rho_\star'(1)>1$, so $\rho_\star(z)$ is also the optimal solution of~\eqref{e:varProbFullA} with $\lambda = -\rho_\star'(1)$. Recollecting the re-parametrisation $\alpha = \lambda/(\lambda-1)$ we conclude that the optimal value of the balance parameter $\alpha$, denoted $\alpha_\star$, is given by
\begin{equation}
\label{e:OptAlpha}
\alpha_\star = \frac{\rho_\star'(1)}{\rho_\star'(1)+1}.
\end{equation}
Finally, recalling that~\eqref{e:varProbFullA} was obtained by setting $\beta=\beta_+$ from~\eqref{e:BetaPlusMinus} and that $\alpha_\star\tau_\star'(z)/(\alpha_\star-1)=\rho_\star'(z)$ according to our rescaling, we can apply~\eqref{e:OptAlpha} and~\eqref{e:OptRhoCond1} in succession to conclude that the optimal value of the balance parameter $\beta$ is
\begin{equation}
\label{e:OptBeta}
\beta_\star 
= 1 + (\alpha_\star-1)\normtwo{\rho_\star'+1}
= \frac{\rho_\star'(1) + 1 - \normtwo{\rho_\star'+1}}{\rho_\star'(1)+1} 
= 2.
\end{equation}

\section{Relation to Hagstrom \& Doering's variational problem}
\label{s:HagstromDoering}

The bounding principle formulated by~\citet{Hagstrom2010} can be recovered upon setting $\alpha=2$ and $\beta=2$ in~\eqref{e:varProbFull0}. These values clearly satisfy~\eqref{e:PosAlphaBeta}, and we have seen that the choice $\beta=2$ is optimal. The variational problem for the optimal background field becomes
\begin{equation}
\label{e:HD}
\begin{aligned}
\max_{\tau(z)} \quad & - \normtwo{\tau'}^2 - 2\,\tau(1),\\
\text{subject to} \quad 
&\langle \abs{\nabla\theta}^2 + 2\,\tau'\,w\,\theta \rangle \geq 0 \quad \forall \text{ admissible } \theta,\,w,\\
&\tau(0)=0.
\end{aligned}
\end{equation}
Strictly speaking we should also enforce the boundary condition $\tau'(1)=-1$, but this does not limit the choice of $\tau$ for the same reasons discussed at the end of \S\ref{s:optimizeBP}. 

To bring~\eqref{e:HD} in contact with~\eqref{e:varProbFullC}, we change variables to
$\varphi = 2\tau$ and use the boundary condition $\varphi(0)=0$ to rewrite~\eqref{e:HD} as
%
\begin{equation}
\label{e:HD2}
\begin{aligned}
\max_{\varphi(z)} \quad &\frac{ 1- \normtwo{\varphi'+1}^2 - 2\,\varphi(1) }{4},\\
\text{subject to} \quad 
&\langle \abs{\nabla\theta}^2 +\varphi'\,w\,\theta \rangle \geq 0 \quad \forall \text{ admissible } \theta,\,w,\\
&\varphi(0)=0.
\end{aligned}
\end{equation}

It is clear that~\eqref{e:varProbFullC} and~\eqref{e:HD2} have the same feasible set. It is also not difficult to show that the optimal value of~\eqref{e:varProbFullC} is no smaller than that of~\eqref{e:HD2}; 
in fact, for any feasible $\varphi(z)$
\begin{equation}
\frac{1 - \normtwo{{\varphi}'+1} - \varphi(1)}{2}
-\frac{ 1- \normtwo{{\varphi}'+1}^2 - 2\,\varphi(1) }{4} 
=
\left( \frac{1 - \normtwo{{\varphi}'+1}}{2} \right)^2
\geq 0.
\end{equation}
In particular, using~\eqref{e:varProbFullC} it is almost immediate to obtain a 4.2\% improvement for the prefactor of Hagstrom \& Doering's bound, $\Nu\leq 0.838\,\Ma^{2/7}$, at least in the limit of infinite Marangoni number: in appendix~\ref{a:HDproof} we show that
\begin{equation}
\Nu \leq 0.803\times \Ma^{2/7}
\quad
\text{as } \Ma\to\infty.
\end{equation}



What is not immediately apparent when comparing~\eqref{e:HD2} to~\eqref{e:varProbFullC} is that fixing the balance parameters a priori does not change the asymptotic behaviour of the optimal bounds as $\Ma\to\infty$. To show that this is true, recall from \S\ref{s:optimalBeta} that the choice $\beta=2$ is optimal. After fixing $\beta=2$ and re-parametrising $\alpha=\lambda/(\lambda-1)$ as in \S\ref{s:BackgroundMethod}---with $\lambda>1$ to satisfy~\eqref{e:PosAlphaBeta}---the bound in~\eqref{e:NuBound2} becomes
\begin{equation}
\label{e:NuBoundBeta1}
\frac{1}{\Nu} \geq 1 - \frac{\lambda^2}{4(\lambda-1)}\normtwo{\tau'+1}^2.
\end{equation}
From this point onwards, the analysis is analogous to that of the infinite-{\Pran} {\RB} problem~\cite[Chapter 6]{Plasting2004}. First, let $w=\Ma\,\tilde{w}$ and define the scaled Marangoni number $M=\lambda\Ma$ to rewrite the spectral constraint~\eqref{e:SClambda} as
\begin{equation}
\label{e:SCscaled}
\left\langle \vert \nabla\theta\vert ^2 + M\,\tau'\,\tilde{w}\,\theta\right\rangle \geq 0
\quad \forall \text{ admissible } \theta,\,\tilde{w}.
\end{equation}
Upon rescaling $w=\Ma\,\tilde{w}$ the Marangoni number drops out of equation~\eqref{e:wDef_theta}, so $\tilde{w}$ is a (linear) function of $\theta$ only and the admissible test functions in~\eqref{e:SCscaled} are independent of $M$. 
%
%
Then, consider the family of background fields $\tau_M$, parametrized by the scaled Marangoni number $M$, that maximises the right-hand side of~\eqref{e:NuBoundBeta1} for a fixed value of $\lambda$. In other words, assume that $\tau_M$ solves the variational problem
\begin{equation}
\begin{aligned}
\min_{\tau(z)} \quad&\normtwo{\tau'+1}^2
\\
\text{subject to} \quad &\left\langle \vert \nabla\theta\vert ^2 + M\,\tau'\,\tilde{w}\,\theta\right\rangle \geq 0
\quad \forall \text{ admissible } \theta,\,\tilde{w}.
\end{aligned}
\end{equation}
Moreover, suppose $\sigma(M)$ is such that
\begin{equation}
\normtwo{\tau_M'+1}^2 = 1 - \sigma(M).
\end{equation}
Note that it is reasonable to assume that $\sigma(M)\to 0$ as $M\to \infty$, because we expect that $\tau'(z)\approx 0$ except for thin boundary layers if the scaled spectral constraint~\eqref{e:SCscaled} is to be satisfied. The optimal bound for a given value of $\lambda$ is then given by
\begin{equation}
\frac{1}{\Nu} \geq \frac{\lambda^2\,\sigma(M) - (\lambda-2)^2}{4(\lambda-1)}.
\end{equation}
%
Using the fact that 
${\rm d}M/{\rm d}\lambda={\rm d}(\lambda\,\Ma)/{\rm d}\lambda=M/\lambda$,
it is straightforward to show that the right-hand side of the last expression is maximised with respect to $\lambda$ when
\begin{equation}
\label{e:optLambda}
\lambda = 
\frac{2 - 2\,\sigma(M) - M\,\sigma'(M)}{1 - \sigma(M)-M\,\sigma'(M)},
\end{equation}
and that the corresponding bound on the Nusselt number is
\begin{equation}
\Nu \leq 
\frac{4 - 4\,\sigma(M) - 4\,M\,\sigma'(M)}
{4\,\sigma(M) - \left[ 2\,\sigma(M)+M\,\sigma'(M)\right]^2}.
\end{equation}
Now, recall that $\Nu\geq 1$ since convection enhances the purely conductive vertical heat transport. Using the fact that $\lambda>1$ and the assumption that $\sigma(M)\to 0$ as $M\to 0$ it is then not difficult to see that since $\Nu\geq 1$ the quantity 
$M\,\sigma'(M)$ 
must be uniformly bounded as the scaled Marangoni number $M$ tends to infinity. Consequently, the solution $\lambda_\star=\lambda_\star(M)$ of~\eqref{e:optLambda} satisfies
\begin{equation}
\lim_{M\to \infty} \lambda_\star(M) = O(1).
\end{equation}
This implies that $\Ma = O(M)$ as $M\to\infty$, meaning that the optimization over the balance parameter does not influence the asymptotic scaling of the bound on $\Nu$ with the Marangoni number. 
\section{Optimal bounds}
\label{s:numerics}

We now turn our attention to the numerical solution of the variational problem~\eqref{e:varProbFullC}. To implement our computational strategy, described in \S\ref{s:method} below, it is convenient  to change variables once more and let
\begin{equation}
\label{e:ChangeVariables}
\rho(z) \defeq \int_0^z \left[ \phi(\xi) - 1 \right] \dxi,
\end{equation}
so the boundary condition $\rho(0)=0$ is satisfied. Since $\rho'(z)=\phi(z)-1$, \eqref{e:varProbFullC} can be rewritten as
\begin{equation}
\label{e:compVP1}
\begin{aligned}
\max_{\phi(z)} \quad & 1 - \frac{1}{2}\normtwo{\phi} 
- \frac{1}{2}\int_0^1 \phi(z),\\
\text{subject to} \quad 
&\langle \abs{\nabla\theta}^2 + (\phi-1)\,w\,\theta \rangle \geq 0 \quad \forall \text{ admissible } \theta,\,w.\\
\end{aligned}
\end{equation}
Moreover, we introduce a non-negative variable $s$ such that $\normtwo{\phi}\leq s$. After dropping the constant $1$ as well as a factor of $1/2$ from the objective function, it is not difficult to see that the optimal solution of~\eqref{e:compVP1} is the same as that of the convex problem
\begin{equation}
\label{e:compVP2}
\begin{aligned}
\max_{\phi(z),\,s} \quad & -s 
- \int_0^1 \phi(z),\\
\text{subject to} \quad 
&\langle \abs{\nabla\theta}^2 + (\phi-1)\,w\,\theta \rangle \geq 0 \quad \forall \text{ admissible } \theta,\,w,\\
&\normtwo{\phi} \leq s.
\end{aligned}
\end{equation}
%

As anticipated in \S\ref{s:introduction}, we are also interested in optimising
the bound on {\Nu} over the restricted classes of monotonically decreasing and convex scaled background fields, {\ie} such that $\rho'(z) \leq 0$ and $\rho''(z)\geq 0$. This is achieved by solving the convex problems
\begin{equation}
\label{e:compVP3}
\begin{aligned}
\max_{\phi(z),\,s} \quad & -s 
- \int_0^1 \phi(z),\\
\text{subject to} \quad 
&\langle \abs{\nabla\theta}^2 + (\phi-1)\,w\,\theta \rangle \geq 0 \quad \forall \text{ admissible } \theta,\,w,\\
&\normtwo{\phi} \leq s,\\
&\phi(z) \leq 1,
\end{aligned}
\end{equation} 
and
\begin{equation}
\label{e:compVP4}
\begin{aligned}
\max_{\phi(z),\,s} \quad & -s 
- \int_0^1 \phi(z),\\
\text{subject to} \quad 
&\langle \abs{\nabla\theta}^2 + (\phi-1)\,w\,\theta \rangle \geq 0 \quad \forall \text{ admissible } \theta,\,w,\\
&\normtwo{\phi} \leq s,\\
&\phi'(z) \geq 0.
\end{aligned}
\end{equation} 
\subsection{Computational methodology}
\label{s:method}


Our computational methodology is based on the observation that the constraints in~\eqref{e:compVP2}--\eqref{e:compVP4} are the infinite-dimensional equivalent of well-known types of finite-dimensional convex constraints. 
As already pointed out in previous work~\citep{Fantuzzi2016PRE} the spectral constraint is the infinite-dimensional equivalent of a {\it linear matrix inequality} (LMI), the condition that a symmetric matrix $\mat{S}$ whose entries are affine with respect to a set of optimisation variables is positive semidefinite (denoted by $\mat{S}\succeq 0$). 
The norm constraint $\normtwo{\phi}\leq s$, instead, is the infinite-dimensional version of a {\it second order cone constraint} (SOCC), {\ie} the requirement that a vector $\vec{y}\in\mathbb{R}^{n+1}$ and a scalar $s$ satisfy $\norm{\vec{y}} \leq s$, where $\norm{\cdot}$ denotes the usual Euclidean norm of a vector.
Finally, the pointwise constraints $\phi(z)\leq 1$ and $\phi'(z)\geq 0$ are the infinite-dimensional equivalent of element-wise inequalities for a vector $\vec{y}\in\mathbb{R}^{n+1}$ of the form $\vec{A}\vec{y} \leq \vec{b}$, with $\vec{A}\in\mathbb{R}^{m\times (n+1)}$ and $\vec{b}\in \mathbb{R}^m$ given.
For more details on LMIs and SOCCs we refer the reader to the works by~\citet{Boyd1994} and~\citet{Boyd2004}. Optimisation problems with LMIs, SOCCs, and element-wise vector inequalities are well-known instances of so-called {\it conic programmes}, and can be solved to high accuracy in polynomial time~\citep{Vandenberghe1996,Boyd2004}. Consequently, problems~\eqref{e:compVP2}--\eqref{e:compVP4} can be solved numerically if we discretise them to obtain conic programmes.

\begin{figure}
\centerline{
\includegraphics[scale=1]{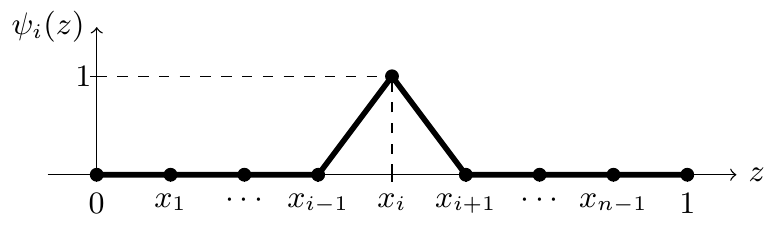}
}
\caption{Sketch of the piecewise-linear function $\psi_i(z)$.}
\label{f:basisFunction}
\end{figure}

In order to reduce the norm constraint $\normtwo{\phi} \leq s$ to a SOCC, we introduce a piecewise-linear ansatz for $\phi$. Given a set of $n+1$ collocation points $0 = z_0 < z_1 < \ldots < z_{n-1} < z_n = 1$, we denote $\phi_i = \phi(z_i)$ for all $i=1,\,\ldots,\,n$ and consider
\begin{equation}
\label{e:FEphi}
\phi(z) = \sum_{i=0}^n \phi_i\,\psi_i(z),
\end{equation}
where $\psi_i(z)$ is the unique piecewise-linear function satisfying $\psi_i(z_i)=1$ and vanishing at all other nodes (cf. figure~\ref{f:basisFunction}). After defining the column vector of nodal values
\begin{equation}
\vec{\Phi} \defeq \left[ \phi_0,\,\ldots,\, \phi_n\right]^T \in \mathbb{R}^{n+1},
\end{equation}
it is clear that there exists a positive definite matrix 
$\mat{P}= \mat{R}^T\mat{R}$
such that
\begin{equation}
\normtwo{\phi} = 
\left(\int_0^1 \sum_{i,j=0}^n \phi_i\,\phi_j\,\psi_i(z)\,\psi_j(z)\dz\right)^{1/2}
= \left( \vec{\Phi}^T\mat{P}\,\vec{\Phi} \right)^{1/2}
= \norm{\mat{R}\,\vec{\Phi}}.
\end{equation}
%
The norm constraint $\normtwo{\phi}\leq s$ then becomes the SOCC
%
$\norm{\mat{R}\,\vec{\Phi}} \leq s$.

The spectral constraint can be reduced to a set of LMIs in a similar way. Recall from \S\ref{s:BM_BalanceParameters} that the spectral constraint is equivalent to the functional $\mathcal{Q}_k\{\hat{\theta}_k\}$ in~\eqref{e:QkDef} being positive semidefinite for all wavenumbers $k\geq 1$ and all complex-valued functions $\hat{\theta}_k(z)$ satisfying $\hat{\theta}_k(0)=0=\hat{\theta}_k'(1)$. Recognising that the real and imaginary parts of $\hat{\theta}_k$ give identical and independent contributions to $\mathcal{Q}_k\{\hat{\theta}_k\}$, it suffices to restrict our attention to real-valued functions $\hat{\theta}_k(z)$, so we define the space of test functions
\begin{equation}
\Gamma \defeq \left\{ 
v(z): [0,1]\to\mathbb{R},\,
\int_0^1 \left( \abs{v'(z)}^2 + \abs{v(z)}^2 \right)\!\dz < \infty,\,
v(0)=0,\,v'(1)=0
\right\}.
\end{equation}
Recalling that we have changed variables according to
\begin{equation}
\frac{\alpha}{\alpha-1}\,\tau'(z) = \rho'(z) = \phi(z)-1,
\end{equation}
we can therefore replace the spectral constraint in~\eqref{e:compVP2}--\eqref{e:compVP4} with the infinite set of {\it Fourier-transformed spectral constraints}
\begin{multline}
\label{e:FTSCphi}
\mathcal{Q}_k\{v\}\!=\!
\int_0^1\!\left\{ \abs{v'(z)}^2 +k^2\abs{v(z)}^2
-\Ma\left[\phi(z)-1\right]f_k(z)\,v(1)\,v(z)
\right\}
\!\dz \geq 0 
\\
\forall v\in\Gamma, \, k=1,\,2,\,\ldots.
\end{multline}
In Appendix~\ref{ss:kCritBound} we show that $\mathcal{Q}_k\{v\}$ is positive semidefinite for a candidate $\phi(z)$ whenever
\begin{equation}
\label{e:kc}
k > k_c \defeq 
\left\lfloor 
\left(\frac{3\sqrt{3}}{128}\right)^{1/4} \Ma^{1/2} \norm{\phi-1}_\infty^{1/2} 
\right\rfloor,
\end{equation}
where $\lfloor\cdot\rfloor$ denotes the integer part of a number. The ``cut-off'' wavenumber $k_c$ represents an upper bound on the largest critical wavenumber, {\ie} the largest values of $k$ for which the infimum of the functional $\mathcal{Q}_k$ in~\eqref{e:FTSCphi} over nonzero test functions is zero. When $k\leq k_c$, instead, we approximate the test function $v$ using the same piecewise-linear ansatz used for $\phi$, {\ie}
\begin{equation}
\label{e:FEv}
v(z) = \sum_{i=0}^n v_i\,\psi_i(z).
\end{equation}
We also set $v_0=0$ and $v_{n} = v_{n-1}$ in order to enforce the boundary conditions $v(0)=0$ and $v'(1)=0$, but we do not do this explicitly in~\eqref{e:FEv} to simplify the following discussion. Substituting~\eqref{e:FEv} and~\eqref{e:FEphi} into $\mathcal{Q}_k\{v\}$ from~\eqref{e:FTSCphi} yields
\begin{multline}
\label{e:QkExp}
\mathcal{Q}_k\{v\} =
\sum_{i,j=0}^{n} v_i\,v_j 
\int_0^1 \left[\psi_i(z)\,'\psi_j(z)' + k^2\,\psi_i(z)\,\psi_j(z)\right]\!\dz
\\
+\Ma \sum_{i=0}^n v_n\,v_i \int_0^1 \psi_i(z)\,f_k(z)\dz
-\Ma \sum_{i,j=0}^{n} \phi_i\,v_n\,v_j \int_0^1 \psi_i(z)\,\psi_j(z)\,f_k(z) \dz.
\end{multline}
Recollecting that we have set $v_0=0$ and $v_{n} = v_{n-1}$, the right-hand side of~\eqref{e:QkExp} is a quadratic form of the vector of nodal values $\vec{v}\defeq[v_1,\,\ldots,\,v_{n-1}]^T$, and there exists an $(n-1)\times(n-1)$ symmetric matrix  $\mat{Q}_k(\vec{\Phi})$, affine with respect to $\vec{\Phi}$, such that
%
$\mathcal{Q}_k\{v\} = \vec{v}^T \mat{Q}_k(\vec{\Phi})\vec{v}$.
%
Consequently, for each wavenumber $k\leq k_c$ the Fourier-transformed spectral constraint $\mathcal{Q}_k\{v\}\geq 0$ can be approximated by the LMI $\mat{Q}_k(\vec{\Phi}) \succeq 0$.

Finally, it is easy to see that the piecewise-linear approximation~\eqref{e:FEphi} turns the pointwise inequality $\phi(z)\leq 1$ into the $n+1$ constraints $\phi_i\leq 1$, $i=0,\,\ldots,\,n$, which can be  written succinctly as the element-wise vector inequality
%
$\vec{\Phi} \leq \vec{1}$.
%
Similarly, the condition $\phi'(z)\geq 0$ becomes a set of $n$ inequalities $\phi_{i-1} - \phi_i \leq 0$, $i=1,\,\ldots,\,n$, which can be written in the vector form
\begin{equation}
\vec{A} \vec{\Phi} \leq \vec{0},\qquad 
\vec{A} \defeq \begin{bmatrix}
1 & -1 \\ & \ddots & \ddots \\ & & 1 & -1
\end{bmatrix} \in \mathbb{R}^{n\times(n+1)}.
\end{equation}

After substituting~\eqref{e:FEphi} into the objective function of~\eqref{e:compVP2} and defining
\begin{equation}
\vec{c} \defeq \left[ \int_0^1 \psi_0(z),\,\ldots,\,\int_0^1 \psi_n(z) \dz\right]^T,
\end{equation}
we can therefore approximate the infinite-dimensional variational problem~\eqref{e:compVP2} with the finite-dimensional conic programme
\begin{equation}
\label{e:CP}
\begin{aligned}
\max_{s,\vec{\Phi}} \quad &-s-\vec{c}^T\vec{\Phi}
\\
\text{subject to} \quad 
&\mat{Q}_k(\vec{\Phi}) \succeq 0, &k=1,\,\ldots,\, k_c,\\
&\norm{\mat{R}\vec{\Phi}} \leq s.
\end{aligned}
\end{equation}
Similarly,~\eqref{e:compVP3} can be approximated as
\begin{equation}
\label{e:CP2}
\begin{aligned}
\max_{s,\vec{\Phi}} \quad &-s-\vec{c}^T\vec{\Phi}
\\
\text{subject to} \quad 
&\mat{Q}_k(\vec{\Phi}) \succeq 0, &k=1,\,\ldots,\, k_c,\\
&\norm{\mat{R}\vec{\Phi}} \leq s,\\
&\vec{\Phi}\leq \vec{1},
\end{aligned}
\end{equation}
while~\eqref{e:compVP4} becomes
\begin{equation}
\label{e:CP3}
\begin{aligned}
\max_{s,\vec{\Phi}} \quad &-s-\vec{c}^T\vec{\Phi}
\\
\text{subject to} \quad 
&\mat{Q}_k(\vec{\Phi}) \succeq 0, &k=1,\,\ldots,\, k_c,\\
&\norm{\mat{R}\vec{\Phi}} \leq s,\\
&\vec{A}\vec{\Phi}\leq \vec{0}.
\end{aligned}
\end{equation}

Before describing our numerical implementation of~\eqref{e:CP}--\eqref{e:CP3} in more detail, let us remark some important aspects of our finite-dimensional approximations. 

The first observation is that introducing the piecewise-linear ansatz~\eqref{e:FEphi} for $\phi(z)$ means that only lower bounds on the optimal values of~\eqref{e:compVP2}--\eqref{e:compVP4} can be computed, because the ``true'' optimal $\phi(z)$ is unlikely to be piecewise-linear. Moreover, assuming~\eqref{e:FEv} enforces the Fourier-transformed spectral constraint only over a particular subset of the test function space $\Gamma$. This enlarges the set of feasible functions $\phi(z)$, so~\eqref{e:CP}--\eqref{e:CP3} yield upper limits for lower bounds of the true optimal values of~\eqref{e:compVP2}--\eqref{e:compVP4}, respectively.  However, one expects the solutions of each conic programme~\eqref{e:CP}--\eqref{e:CP3} to converge to that of the corresponding maximisation problem~\eqref{e:compVP2}--\eqref{e:compVP4} as the number of collocation points in the spatial discretisation increases.

One could also estimate the error between functions in $\Gamma$ and their finite-dimensional approximation, in order to formulate conic programmes whose optimal solutions bound the optimal value of~\eqref{e:compVP2}-~\eqref{e:compVP4} rigorously from below. This is possible if global polynomial approximation is utilised when all but the test functions in the spectral constraint are polynomials~\citep{Fantuzzi2017TAC}. One could follow a similar line of reasoning in each sub-interval of our piecewise-linear approximation, with the additional complication that the function $f_k$ appearing in the spectral constraint is not polynomial. However, we do not do so here because we do not aim to compute bounds on the Nusselt number to the standard of a computer-assisted proof. 


Finally, as already pointed out in \S\ref{s:introduction}, a major advantage of our computational methodology is that the monotonicity and convexity constraints can be implemented in a very straightforward way. On the contrary, optimising the bound on {\Nu} over all monotonic or convex background fields seems considerably more challenging if one follows the classical Euler--Lagrange variational approach, because one has to solve a set of differential equations coupled to an inequality (in fact, a {\it differential} inequality in the convex case).



\subsection{Implementation details}
\label{s:implementation}

The conic programmes~\eqref{e:CP}--\eqref{e:CP3} 
were set up using the MATLAB toolbox YALMIP~\citep{Lofberg2004} and solved with the conic solver SDPT3~\citep{Toh1999,Tutuncu2003}. Sparsity was exploited using chordal decomposition methods~\citep{Fukuda2000,Nakata2003,Kim2011}. All computations were run on a PC with a 3.40GHz Intel\textsuperscript{\textregistered} Core{\texttrademark} i7-4770 CPU and 16Gb of RAM.

As collocation points, we used the Chebyshev nodes $z_i = [1-\cos(\pi i/n)]/2$, $i=0,\,\ldots,\,n$  in the sub-interval $(0.05,0.98)$, and the finer distribution $z_i = [1-\cos(\pi i/4n)]/2$, $i=0,\,\ldots,\,4n$ in the boundary sub-intervals $[0,0.05]$ and $[0.98,1]$. After initial experiments we set $n=512$, giving $873$ collocation points in total; our results, presented in \S\ref{s:results}, change by less that 0.1\% if a larger $n$ is used.

Chebyshev nodes were chosen as they naturally cluster near the boundaries and help resolve boundary layers near $z=0$ and $z=1$ in the optimal $\phi(z)$.  These are expected even if no boundary conditions are imposed because to maximise the objective function in~\eqref{e:compVP2} one would like to choose $\phi(z)<0$, but setting $\phi(z)\approx 1$ in the bulk of the domain is necessary to be able to satisfy the spectral constraint. However, it is possible to have $\phi(z)<0$ in thin layers near the walls because the functions $f_k$, which act as a weight on $\phi$ in the Fourier-transformed spectral constraint~\eqref{e:FTSCphi}, are small there for all $k$'s (cf. figure~\ref{f:fkfigure}). These observations are confirmed by the numerical results presented in \S\ref{s:results}.

While boundary layers can in principle be resolved with a sufficiently fine distribution of Chebyshev points, we preferred to refine the discretisation only near the boundaries using a secondary set of Chebyshev nodes to limit the cost of our computations. 
%
A precise assessment of the computational burden of our conic programmes relies on technical details of the sparsity-exploiting methods we used~\citep{Fukuda2000,Nakata2003,Kim2011} and is beyond the scope of this work. Here, we simply note that it must grow at least linearly with the number of collocation points. Roughly speaking, in fact, sparsity allows replacing each LMI $\mat{Q}_k(\vec{\Phi}) \succeq 0$ with a set of LMIs on certain $3\times 3$ submatrices of $\mat{Q}_k$. Since the number of rows/columns of $\mat{Q}_k$ grows linearly with the number of discretisation points, so does the number of such submatrices. Even assuming optimistically that the computational cost of one such $3\times 3$ LMI is fixed and that handling a much larger number of LMIs has negligible overhead, the overall computational cost can grow no slower than linearly with the number of collocation nodes.

One complication to the implementation of~\eqref{e:CP} is that the cut-off wavenumber $k_c$ is not known a priori, but it depends on $\vec{\Phi}$ according to~\eqref{e:kc}. We therefore employ the following iterative procedure: find the optimal $\vec{\Phi}$ using an initial guess $k_0$ for $k_c$, update the value of $k_c$ using~\eqref{e:kc}, check if $\mat{Q}_k(\vec{\Phi})$ is positive semidefinite for all $k \leq k_c$, and repeat the optimisation with the updated guess for $k_c$ if any of these checks fail.


A second hurdle is that solving~\eqref{e:CP} with this iterative procedure becomes expensive when the Marangoni number is large because the cut-off wavenumber $k_c$, and therefore the number of LMI constraints, grows proportionally to $\Ma^{1/2}$. For example, at $\Ma=2.5\times 10^6$ we find that the optimal $\phi$ satisfies $\norm{\phi-1}_\infty =2$, so~\eqref{e:kc} gives $k_c=1\,003$; when all $1\,003$ LMIs are considered in~\eqref{e:CP}, SDPT3 takes more than 4 hours to converge on our machine. 
In an effort to reduce the CPU time requirements, we implemented a trial-and-error procedure, inspired by the numerical continuation method employed by~\citet{Plasting2003}, in which only a subset of wavenumbers are considered in~\eqref{e:CP}. More precisely, we progressively increased the Marangoni number according to the update rule $\Ma_{i+1}=\Ma_i \times 10^p$, which gives $p+1$ logarithmically spaced points between successive powers of 10. Given the critical wavenumbers $k_1,\,\ldots,\,k_m$ at one Marangoni number, we solved the SDP for the next {\Ma} considering only wavenumbers in a window of width $2r$ around each $k_i$, $i=1,\,\ldots,\,m$, {\ie} values of $k$ such that
\begin{equation}
\label{e:critWavenoRange}
k \in \bigcup_{i=i}^{m} \,[k_i-r,k_i+r].
\end{equation}
We then checked if the optimal solution satisfied $\mat{Q}_k(\vec{\Phi})\succeq 0$ for all remaining wavenumbers up to the cut-off value $k_c$. If any of these checks failed, we added the wavenumber with the largest constraint violation ({\ie} corresponding to the matrix $\mat{Q}_k$ with the most negative eigenvalue) to the list of critical values and repeated the optimisation.


\subsection{Results}
\label{s:results}


The conic programmes~\eqref{e:CP}--\eqref{e:CP3} were successfully solved for Marangoni numbers up to $\Ma=10^9$ using the procedure described in \S\ref{s:implementation} with $p=19$ and $r=10$. In all cases, at each value of $\Ma$ the optimal $\phi(z)$ was used to recover the optimal scaled background field $\rho(z)$ and the corresponding bound on the Nusselt number.

\begin{figure}
\centering
\includegraphics[scale=1]{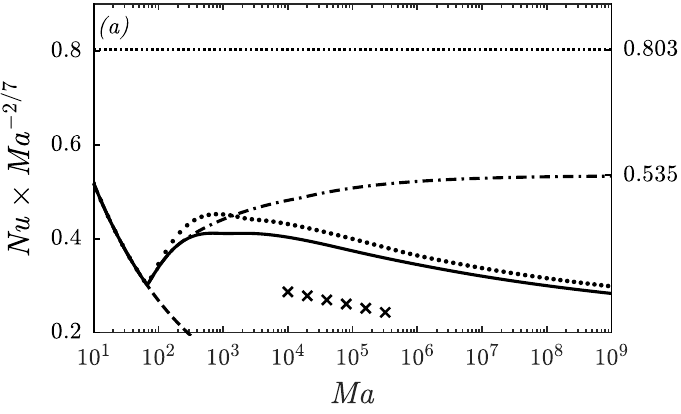}\hfill
\includegraphics[scale=1]{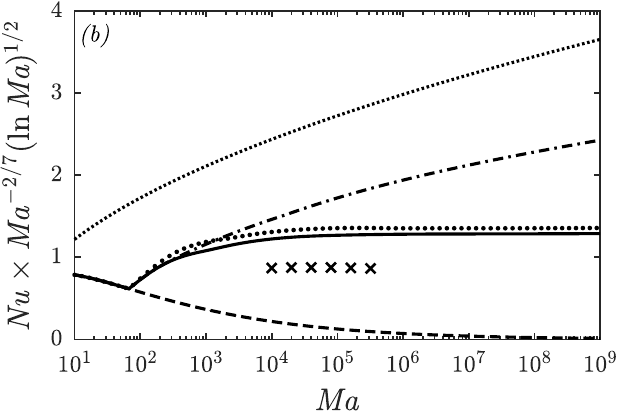}
\caption{Comparison between: the fully optimal bounds on the Nusselt number, computed using the solution of~\eqref{e:compVP2} (solid line); the optimal monotonic bounds, computed using the solution of~\eqref{e:compVP3} (dot-dashed line); the optimal convex bounds, computed using the solution of~\eqref{e:compVP4} (thick dotted line).
Also shown are the conductive Nusselt number $\Nu=1$ (dashed line), the analytical bound $\Nu\leq 0.803\,\Ma^{2/7}$ proven in \S\ref{s:HagstromDoering} (dotted line), and the DNS data~\citep[][crosses]{Boeck2001}. In subfigure {\it(a)}, the data are compensated by $\Ma^{-2/7}$ to facilitate the visual comparison with the asymptotic scaling of the analytical bound. In subfigure {\it(b)}, the data are compensated by $\Ma^{-2/7}(\ln\Ma)^{1/2}$.}
\label{f:bounds}
\end{figure}

The most important results of our computations are the bounds on {\Nu}, which are plotted in figure~\ref{f:bounds}. Also shown for comparison are: the analytical bound $\Nu \leq 0.803\,\Ma^{2/7}$ from \S\ref{s:HagstromDoering}; the DNS results obtained by~\citet{Boeck2001}; the conductive value $\Nu=1$, which bounds the Nusselt number from below. The results are plotted in two ways: compensated by a factor of $\Ma^{-2/7}$ to aid the visual comparison with the asymptotic scaling of the analytical bound, and compensated by $\Ma^{-2/7}(\ln\Ma)^{1/2}$. 

The main observation is that while a gap with the DNS data remains, the fully optimal bounds and those computed after enforcing convexity grow more slowly than the analytical bound by $(\ln\Ma)^{1/2}$. In particular, the fully optimal bounds exhibit the asymptotic behaviour 
\begin{equation}
\label{e:conjecture}
\Nu \leq 1.285 \Ma^{2/7} (\ln \Ma)^{-1/2}.
\end{equation}
In contrast, when the background field is constrained to decrease monotonically the bound on {\Nu} asymptotes to $0.535 \Ma^{2/7}$.
This suggests that the analytical bound of \S\ref{s:HagstromDoering} attains the optimal asymptotic scaling available when $\rho(z)$ is monotonic, but it may be lowered by a logarithm upon construction of a non-monotonic background field.

\begin{figure}
\centering
\includegraphics[scale=0.95]{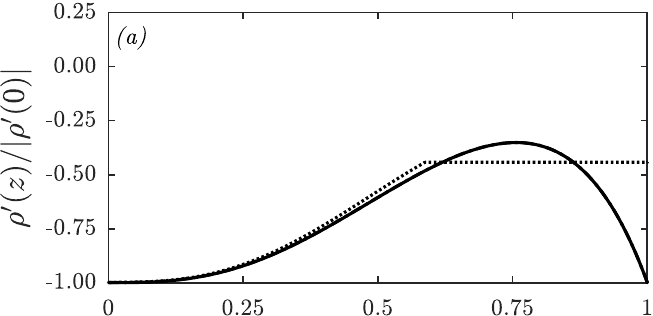}\hspace{1em}
\includegraphics[scale=0.95]{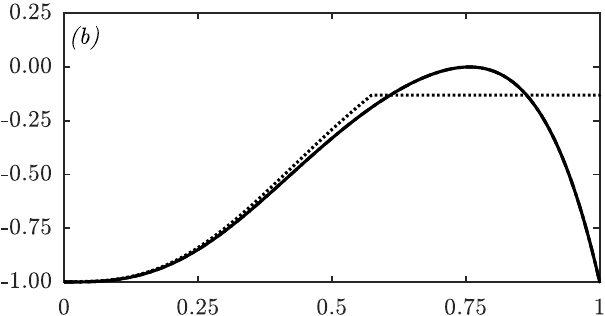}\\[0.5em]
\includegraphics[scale=0.95]{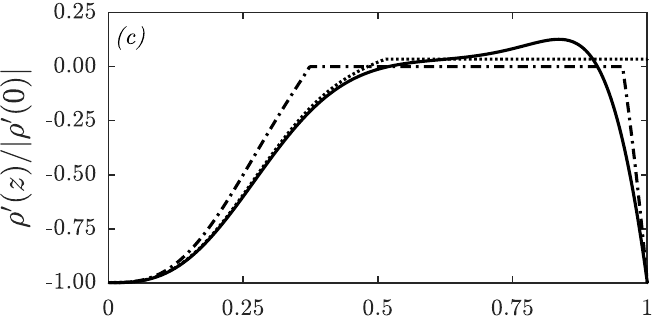}\hspace{1em}
\includegraphics[scale=0.95]{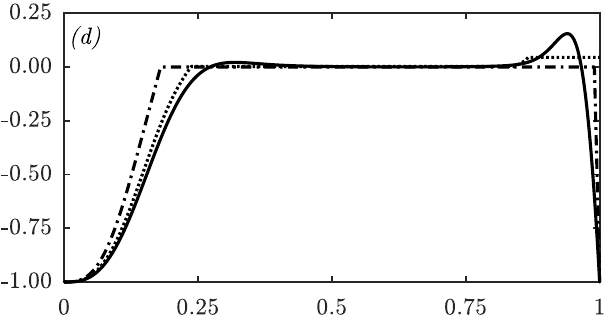}\\[0.5em]
\includegraphics[scale=0.95]{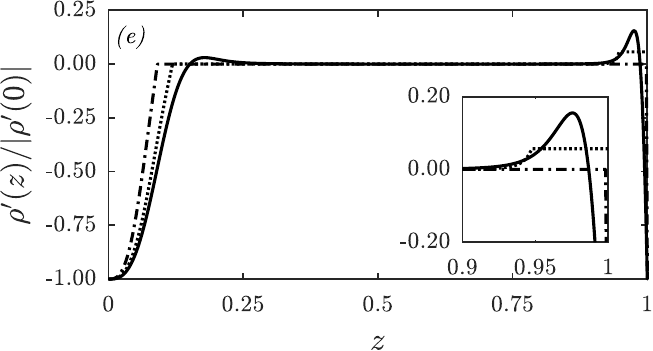}\hspace{1em}
\includegraphics[scale=0.95]{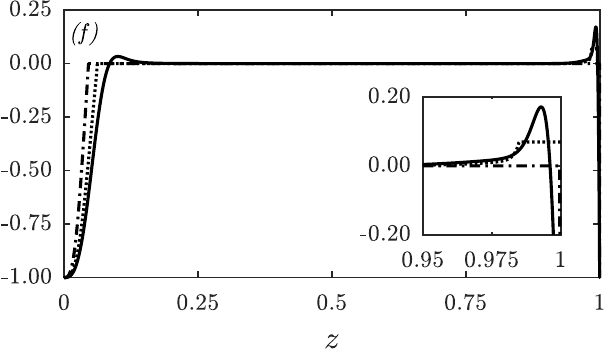}
\caption{Normalised derivatives of the optimal background fields, $\rho'(z)/\abs{\rho'(0)}$, obtained with~\eqref{e:CP} (solid line), with~\eqref{e:CP2} (dot-dashed lines), and~\eqref{e:CP3} (dotted line) for: {\it(a)} $\Ma = 100$;
{\it(b)} $\Ma = 186.12$;
{\it(c)} $\Ma = 10^3$;
{\it(d)} $\Ma = 10^4$;
{\it(e)} $\Ma = 10^5$; and
{\it(f)} $\Ma = 10^6$.
Inserts in {\it(e)} and {\it(f)} show a detailed view of the boundary layers near $z=1$.}
\label{f:BF}
\end{figure}

\begin{figure}
\centering
\includegraphics[scale=1]{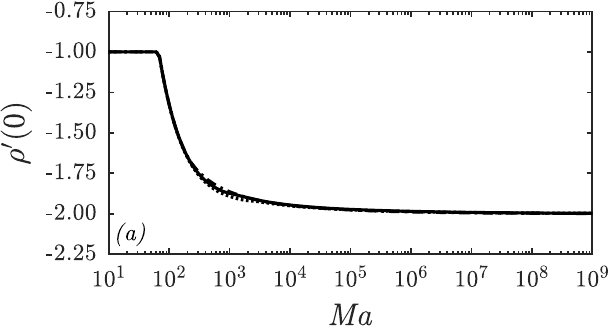}\hspace{1em}
\includegraphics[scale=1]{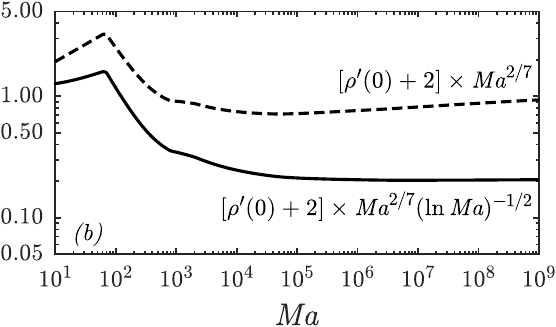}
\caption{
{\it (a)}
Boundary value $\rho'(0)$ for the fully optimal (solid line), monotonic (dot-dashed line), and convex (dotted line) background fields. All curves almost coincide.
{\it (b)}
Plot of the convergence measure $\rho'(0)+2$, scaled by $\Ma^{2/7}(\ln \Ma)^{-1/2}$ (solid line) and by $\Ma^{2/7}$ (dashed line), for the fully optimal background fields.
}
\label{f:rhoBV}
\end{figure}


Figure~\ref{f:BF} shows the derivative of the optimal scaled background field, computed with each of the conic programmes~\eqref{e:CP}--\eqref{e:CP3}, for selected values of {\Ma}. We plot $\rho'(z)$ instead of $\rho(z)$ because by virtue of~\eqref{e:ChangeVariables} problems~\eqref{e:compVP2}--\eqref{e:compVP4} can be rewritten in terms of $\rho'(z)$ alone. Since $\rho(z)$ can be recovered by integration using the boundary condition $\rho(0)=0$, the derivative $\rho'(z)$ is the actual decision variable in \eqref{e:compVP2}--\eqref{e:compVP4}. Moreover, to ease the comparison the profiles have been normalised by the magnitude of the boundary value $\rho'(0)$, which converges to $-2$ as {\Ma} grows as illustrated in figure~\ref{f:rhoBV}{\it(a)}. Figure~\ref{f:rhoBV}{\it(b)} demonstrates that in the fully optimal case  the convergence is logarithmic; this was also observed when convexity was imposed, while power-law convergence was observed for the monotonic profiles (these results are not show for brevity). Such evidence corroborates our conjecture that~\eqref{e:conjecture} is the correct functional form the optimal bound on {\Nu}. 

As illustrated by figure~\ref{f:BF}, the optimal $\rho'(z)$ is negative for $\Ma\leq 186.12$, meaning that the corresponding scaled background field decreases monotonically for sufficiently small Marangoni numbers. When {\Ma} is raised, all profiles are characterised by boundary layers separated by a bulk region where $\rho'(z)\approx 0$. Note that the transition to the bulk region is not smooth when monotonicity or convexity are enforced, and this is the main reason for preferring the piecewise-linear approximations of \S\ref{s:method} to the global polynomial approximation used in previous works~\citep{Fantuzzi2016PRE,Fantuzzi2016CDC,Fantuzzi2017TAC}. 

In the fully optimal case, $\rho'(z)$ changes sign inside both boundary layers to reach positive local maxima, so the corresponding scaled background field is characterised by non-monotonic boundary layers. Enforcing monotonicity removes these local maxima and makes the boundary layers thinner, while convexity prevents the local maximum near $z=0$ and makes $\rho'(z)$ constant across the boundary layer near $z=1$.

\begin{figure}
\centering
\includegraphics[scale=1]{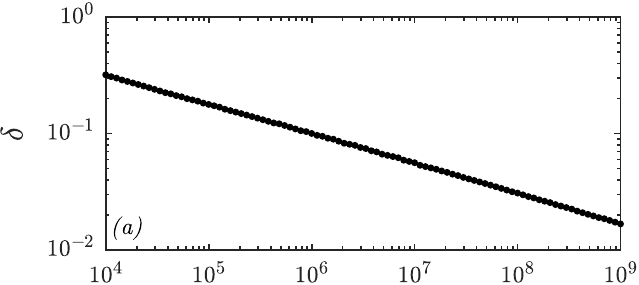}\hspace{0.8em}
\includegraphics[scale=1]{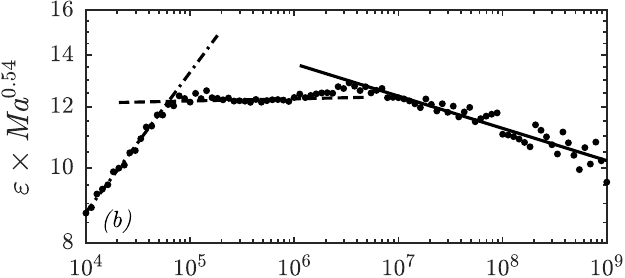}\\[2ex]
\hspace{0.15em}
\includegraphics[scale=1]{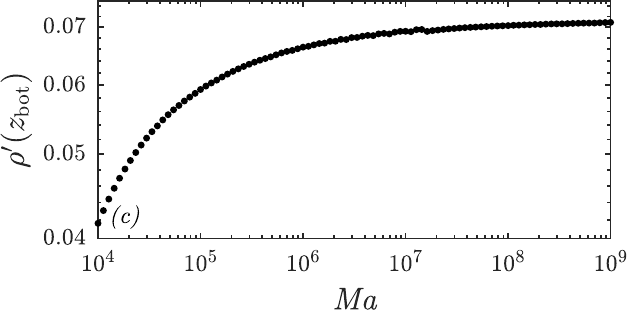}\hspace{0.4em}
\includegraphics[scale=1]{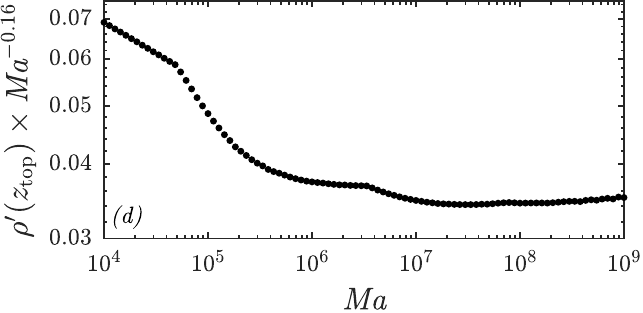}
\caption{
Details of the boundary layer structure of the fully optimal scaled background field derivative $\rho'(z)$ for $\Ma \geq 10^4$. The dot-dashed, dashed, and solid lines in {\it(b)} indicate the approximate scaling laws~\eqref{e:epsScaling1}--\eqref{e:epsScaling3}, respectively.
}
\label{f:BLdataFull}
\end{figure}

Further details of the boundary layer structure of the fully optimal profiles for $\Ma\geq 10^4$ are given in figure~\ref{f:BLdataFull} (very similar results for the optimal convex profiles are not shown for brevity). Letting $z_{\rm bot}$ and $z_{\rm top}$ denote the coordinates of the positive local maxima of $\rho'(z)$ near $z=0$ and $z=1$, respectively, we take $\delta \defeq z_{\rm bot}$ and $\varepsilon \defeq 1-z_{\rm top}$ as measure of the thickness of each boundary layer. The boundary layer near the bottom of the domain ($z=0$) becomes approximately self-similar at large Marangoni numbers, and least-squares power-law fits to the data in figures~\ref{f:BLdataFull}{\it(a)} and~\ref{f:BLdataFull}{\it(c)} for $\Ma\geq 10^7$ return
\begin{align}
\delta &\approx 3.8\,\Ma^{-0.26}, &
\rho'(z_{\rm bot}) &\approx 0.07.
\end{align}
Note that the scaling exponent of $\delta$ is not far from $-2/7 \approx -0.286$, suggesting that the width of the boundary layer near $z=0$ is one of the leading factors determining the scaling of the bound on {\Nu}. In fact, we conjecture that asymptotically $\delta = O(\Ma^{-2/7} (\ln \Ma)^{1/2})$, such that $\Nu = O(\delta^{-1})$, but unfortunately the finite precision of our data does not permit to clearly identify logarithmic corrections. To obtain more precise values we should solve the conic programmes~\eqref{e:CP}--\eqref{e:CP2} to a level of accuracy beyond the capabilities of SDPT3, as well as study larger Marangoni numbers (this issue will be discussed further in \S\ref{s:discussion}).

The situation is more complicated for the boundary layer near $z=1$. In figure~\ref{f:BLdataFull}{\it(b)} we can identify three distinct regions characterised by different scaling laws for $\varepsilon$:
\begin{subequations}
\begin{align}
\label{e:epsScaling1}
\varepsilon &\approx 1.65\,\Ma^{-0.36} 
\qquad\text{for } \Ma\lessapprox 5\times 10^4,\\
\label{e:epsScaling2}
\varepsilon &\approx 11.8\,\Ma^{-0.54} 
\qquad\text{for } 5\times 10^4 \lessapprox \Ma\lessapprox 3\times 10^6,\\
\label{e:epsScaling3}
\varepsilon &\approx 24.3\,\Ma^{-0.58} 
\qquad\text{for } \Ma\gtrapprox 3\times 10^6.
\end{align}
\end{subequations}
In the first and third regions we could also determine approximate scaling laws for the peak value $\rho'(z_{\rm top})$:
\begin{subequations}
\begin{align}
\rho'(z_{\rm top}) &\approx 0.34\,\Ma^{-0.01} 
\qquad\text{for } \Ma\lessapprox 5\times 10^4,\\
\rho'(z_{\rm top}) &\approx 0.04\,\Ma^{0.16} 
\qquad\;\;\,\text{for } \Ma\gtrapprox 3\times 10^6.
\end{align}
\end{subequations}
Once again, these scaling laws are only tentative due to the finite precision to which the conic programmes for the optimal bounds could be solved.  However, we remark that the large scatter in the the data points in figure~\ref{f:BLdataFull}{\it(b)} is simply due to plotting $\varepsilon$ after rescaling by $\Ma^{0.54}$, which at large $\Ma$ amplifies small inaccuracies in our numerical data.

\begin{figure}
\centering
\includegraphics[scale=1]{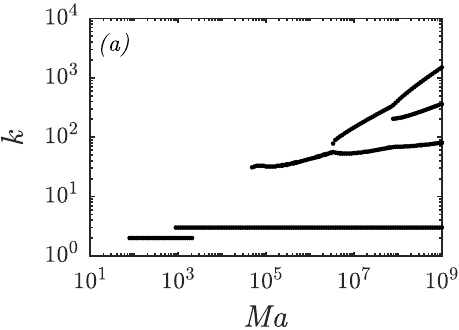}\hfill
\includegraphics[scale=1]{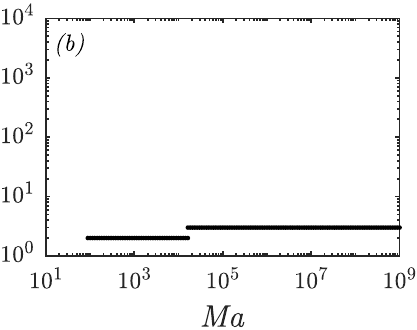}\hfill
\includegraphics[scale=1]{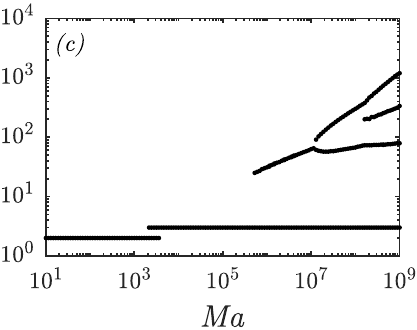}
\caption{
Bifurcation diagrams for the critical wavenumbers in: {\it (a)} the conic programme~\eqref{e:CP} for the fully optimal background fields; {\it (b)} the conic programme~\eqref{e:CP2} for the optimal monotonic background fields; {\it (c)} the conic programme~\eqref{e:CP3} for the optimal convex background fields.
}
\label{f:bifurcations}
\end{figure}

Changes in the scaling of the boundary layer near $z=1$ correspond to bifurcations in the critical wavenumbers for the conic programme~\eqref{e:CP}. As illustrated in figure~\ref{f:bifurcations}{\it(a)}, new critical wavenumbers appear at large values of $k$ for $\Ma\approx 4\times 10^5$ and $\Ma \approx 3\times 10^6$. Another intermediate branch of critical wavenumbers appears for $\Ma \approx 10^8$, but this does not seem to influence the scaling of the boundary layer. Such bifurcations can be explained in terms of the interactions in the Fourier-transformed spectral constraint~\eqref{e:FTSCphi} between the boundary layer of $\rho'(z)=\phi(z)-1$ and the function $f_k(z)$, which is almost entirely supported near $z=1$ at large $k$. As shown in figures~\ref{f:bifurcations}{\it(b)}--{\it(c)}, similar bifurcations were observed when solving~\eqref{e:CP3} but not when solving~\eqref{e:CP2}, probably because the boundary layer near $z=1$ of the optimal monotonic background fields is too thin to allow interesting interactions for wavenumbers below the cut-off value $k_c$.
%

\begin{figure}
\centering
\includegraphics[scale=1]{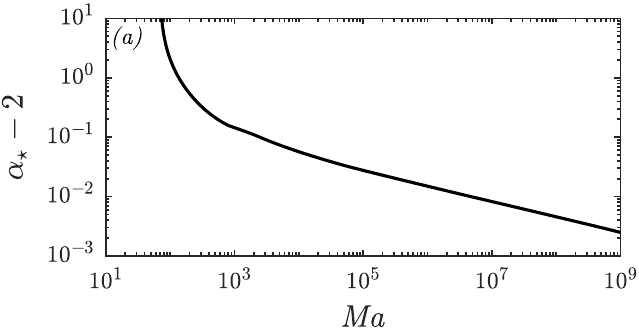} \hspace{1.5em}
\includegraphics[scale=1]{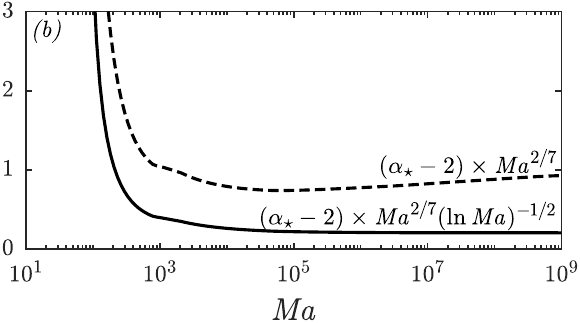}
\caption{
{\it (a)}
Convergence of the optimal balance parameter $\alpha_\star$, computed using the optimal solution of~\eqref{e:compVP2} and~\eqref{e:OptAlpha}, to the asymptotic value $2$.
{\it (b)}
Plot of the difference $\alpha_\star-2$, scaled by $\Ma^{2/7}(\ln\Ma)^{-1/2}$ (solid line) and by $\Ma^{2/7}$ (dashed line).
}
\label{f:alphaOpt}
\end{figure}

To conclude this section, we plot in figure~\ref{f:alphaOpt} the variation with {\Ma} of the optimal balance parameter $\alpha_\star$, computed using~\eqref{e:OptAlpha} and the fully optimal background field. The results are interesting for two reasons. First, the convergence of $\alpha_\star$ to $2$ as {\Ma} is raised is logarithmic, giving further evidence in support of~\eqref{e:conjecture}. Second, the results suggest that the choice $\alpha=2$ in the original work by~\citet{Hagstrom2010}---presumably motivated only by the convenience of eliminating the linear terms when combining~\eqref{e:PertEnEq}, \eqref{e:NuId1}, and \eqref{e:NuId2} in the background method analysis---is optimal, at least as far as the asymptotic behaviour of the bound as $\Ma\to\infty$ is concerned. Contrary to what has been observed in previous works~\citep[see e.g.][]{Plasting2003,Wen2015a}, this means that not only the optimisation of the balance parameters has no influence on the asymptotic scaling of the bound (cf. \S\ref{s:HagstromDoering}), but it also does not improve the optimal prefactor available to \HD's original upper-bounding principle.

\section{Discussion}
\label{s:discussion}


\subsection{Towards an improved bound}
\label{s:improvedbound}
The results presented in \S\ref{s:results} suggest that \HD's bound $\Nu\leq O(\Ma^{2/7})$ may be improved by the logarithmic factor $(\ln\Ma)^{-1/2}$. Despite the strong numerical evidence, however, whether the optimal bound scales logarithmically when $\Ma\to\infty$ remains uncertain due to the limited range of Marangoni numbers spanned the present investigation (see \S\ref{s:scalability} for more on this issue). In particular, we cannot rule out the occurrence of further bifurcations in the critical wavenumbers that may cause a transition to a pure power-law behaviour with scaling exponent of 2/7.

Uncertainty about the true asymptotic scaling notwithstanding, our numerical results demonstrate that if the current analytical bound $\Nu\leq O(\Ma^{2/7})$ can be improved, to do so requires a background temperature profile with non-monotonic boundary layers. 
%
%
More precisely, the optimal convex background fields and the corresponding bounds on {\Nu} are evidence that what is needed is a relatively simple non-monotonic boundary layer near $z=1$, while non-monotonicity near $z=0$ only lowers the prefactor. 

%
%

Taking advantage of these observations to improve the bound on {\Nu} analytically, however, is likely to require a careful analysis of the sign-indefinite term in each Fourier-transformed spectral constraint, which we restate here in terms of the variable $\rho(z)$ in the slightly rearranged form
\begin{equation}
\label{e:FTSC}
\mathcal{Q}_k\{v\}\!=\!
\normtwo{v'}^2+k^2\normtwo{v}^2
-\Ma\,v(1)\int_0^1\rho'(z)\,f_k(z)\,v(z)\dz \geq 0 
\quad \forall v\in\Gamma.
\end{equation}
For example, simply estimating
\begin{equation}
\label{e:RubbishBound}
\abs{ \Ma\,v(1) \int_0^1 \rho'(z)\,f_k(z)\,v(z) \dz} \leq  
\Ma\,\abs{v(1)} \int_0^1 \abs{\rho'(z)}\,\abs{f_k(z)}\,\abs{v(z)} \dz
\end{equation}
and requiring (as we have done in appendix~\ref{a:HDproof}) that
\begin{equation}
\label{e:RubbishEstimate}
\normtwo{v'}^2 + k^2\normtwo{v}^2 
- \Ma\,\abs{v(1)}\int_0^1 \abs{\rho'(z)}\,\abs{f_k(z)}\,\abs{v(z)} \dz
\geq 0
\quad \forall v \in \Gamma,
\end{equation}
 forces the optimal $\rho$ to decrease monotonically. In fact, if $\rho$ satisfies~\eqref{e:RubbishEstimate} and $\rho'(z)\geq 0$ for $z\in\mathcal{U}\subset [0,1]$, the profile
\begin{equation}
\tilde{\rho}'(z) \defeq \begin{cases}
\rho'(z), &z\in[0,1]\smallsetminus\mathcal{U},\\
0,&z\in\mathcal{U},
\end{cases}
\end{equation}
also satisfies~\eqref{e:RubbishEstimate}, but decreases monotonically and gives a larger objective value in~\eqref{e:varProbFullC}. In light of the numerical results presented in \S\ref{s:results}, we expect that any bound obtained using the estimate~\eqref{e:RubbishBound} will not be better than $\Nu\leq O(\Ma^{2/7})$.


A better approach is to reformulate the Fourier-transformed spectral constraint~\eqref{e:FTSC} before applying any estimates. 
Without any loss of generality, let $\delta\in(0,1)$ and write
\begin{equation}
\rho'(z) = \begin{cases}
g(z), &0\leq z \leq \delta,\\
h(z), &\delta \leq z \leq 1.
\end{cases}
\end{equation}
Here, $\delta$ represents the thickness of the boundary layer of the optimal background field near $z=0$.
With this choice, the Fourier-transformed spectral constraint~\eqref{e:FTSC} becomes
\begin{multline}
\mathcal{Q}_k\{v\} = \normtwo{v'}^2 + k^2\normtwo{v}^2 
- \Ma\,v(1)\int_0^\delta g(z)\,f_k(z)\,v(z) \dz
\\
- \Ma\,v(1)\int_\delta^1 h(z)\,f_k(z)\,v(z) \dz \geq 0
\quad \forall v \in \Gamma.
\end{multline}
Since this inequality is homogeneous in $v$ and holds when $v(1)=0$, we may restrict attention to test functions normalised such that $v(1)=1$. Upon adding and subtracting $\Ma\int_\delta^1 h(z)f_k(z) \dz$ we then need to check that
\begin{multline}
\label{e:QkGoodFormulation}
\normtwo{v'}^2 + k^2\normtwo{v}^2 
- \Ma\int_0^\delta g(z)\,f_k(z)\,v(z) \dz
\\
+ \Ma\int_\delta^1 h(z)\,f_k(z)\,[1-v(z)] \dz - \Ma\int_\delta^1 h(z)f_k(z) \dz\geq 0.
\end{multline}
If $\int_\delta^1 h(z)f_k(z) \dz<0$, the last term in~\eqref{e:QkGoodFormulation} gives a net positive contribution to the spectral constraint, and can be used to control the sign-indefinite terms. Recalling from figure~\ref{f:fkfigure} that $f_k(z)\leq 0$, this requires $h(z)>0$ over a sufficient portion of the interval $(\delta,1)$, so the background field $\rho$ does not decrease monotonically. Moreover, $h(z)$ should be supported in a boundary layer near $z=1$ to be able to control the fourth term in~\eqref{e:QkGoodFormulation}. Consequently, a non-monotonic boundary layer near $z=1$ helps enforcing the spectral constraint. The situation is similar in infinite-{\Pran} {\RB} convection~\citep{Doering2006,Otto2011}, so this observation is perhaps not surprising.


In addition to casting light on the role of the surface boundary layer, identity~\eqref{e:QkGoodFormulation} may also offer a starting point to improve the bound $\Nu\leq O(\Ma^{2/7})$ analytically.
Recalling the boundary condition $v(0)=0$ and that $v(1)=1$ by virtue of our choice of normalisation for $v$, one possible approach is to use the fundamental theorem of calculus and the Cauchy--Schwarz inequality to bound
\begin{align}
\abs{\int_0^\delta g(z)\,f_k(z)\,v(z) \dz} 
&\leq \int_0^\delta \abs{g(z)\,f_k(z)}\,\abs{\int_0^z v'(t) \dt} \dz 
\nonumber \\
&\leq \normtwo{v'} \int_0^\delta \abs{g(z)\,f_k(z)}\sqrt{z} \dz
\end{align}
and
\begin{align}
\abs{\int_\delta^1 h(z)\,f_k(z)\,[1-v(z)] \dz} 
&\leq \int_\delta^1 \abs{h(z)\,f_k(z)}\,\abs{\int_z^1 v'(t) \dt} \dz 
\nonumber \\
&\leq \normtwo{v'} \int_\delta^1 \abs{h(z)\,f_k(z)}\sqrt{1-z} \dz.
\end{align}
Defining
\begin{subequations}
\begin{align}
a_k &\defeq \int_0^\delta \abs{g(z)\,f_k(z)}\sqrt{z} \dz,\\
b_k &\defeq \int_\delta^1 \abs{h(z)\,f_k(z)}\sqrt{1-z} \dz,\\
c_k &\defeq - \int_\delta^1 h(z)f_k(z) \dz
\end{align}
\end{subequations}
to ease the notation, a sufficient condition for~\eqref{e:QkGoodFormulation} is that
\begin{equation}
\normtwo{v'}^2 - \Ma\left( a_k + b_k \right) \normtwo{v'} + \Ma\,c_k \geq 0,
\end{equation}
which in turn is satisfied if
\begin{equation}
\label{e:SimpleCondition}
a_k + b_k  \leq 2\,\sqrt{\frac{c_k}{\Ma}}.
\end{equation}
Given a candidate background field, condition~\eqref{e:SimpleCondition} can be checked for all wavenumbers up to the `cut-off' wavenumber $k_c$ in~\eqref{e:kc}.
%

Improving the bound $\Nu\leq O(\Ma^{2/7})$, however, may not be straightforward. To illustrate one of the difficulties, let us consider a simple background field. Motivated by figure~\ref{f:rhoBV}{\it(b)} and the shape of the derivatives of the optimal convex background fields in figure~\eqref{f:BF}, we fix
\begin{align}
g(z) &= -2,
&
h(z) &= \begin{cases}
0,& \delta \leq z < 1-\eps,\\
\gamma, &1-\eps \leq z \leq 1,
\end{cases}
\end{align}
with $\gamma>0$ a constant (independent of the Marangoni number) and $\varepsilon \ll 1$ but such that $1/\eps \leq k_c =O(\Ma^{1/2})$. When $k\leq 1/\eps$ we can use the Taylor expansions $f_k(z) =O(z^2)$ near $z=0$ and $f_k(z) = O(k(z-1))$ near $z=1$ to estimate
\begin{align}
a_k &= O\left(\delta^{7/2}\right), &
b_k &= O\left(\gamma\,k\,\eps^{5/2}\right), &
c_k &= O\left(\gamma\,k\,\eps^2\right).
\end{align}
Using these estimates,~\eqref{e:SimpleCondition} can be rearranged as
\begin{equation}
\label{e:finalbound}
\delta^{7/2} \leq O\left(
2\,\eps\,\sqrt{\frac{\gamma\,k}{\Ma}} \left( 1 - \sqrt{\gamma\,k\,\Ma\,\eps^3}\right)
\right).
\end{equation}
When $k = O(1)$ the two sides of~\eqref{e:finalbound} could be balanced by taking $\varepsilon= O(\Ma^{-1/3})$ and $\delta = O(\Ma^{-5/21})$, and upon computing the bound on {\Nu} we find
\begin{equation}
\Nu \leq \frac{2}{2\delta - \gamma\eps - \sqrt{\gamma(\gamma+2)\eps}} 
= O\left(\frac{1}{\delta}\right)  = O\left(\Ma^{5/21}\right).
\end{equation}
Interestingly, the exponent $5/21 \approx 0.238$ is extremely close to that of the best power-law fit $\Nu = O(\Ma^{0.24})$ to the DNS data by Boeck \& Thess (2001, see equation (4) in their paper). In these simulations convection takes the form of stationary rolls with energy only at low wavenumbers, and the deviation from the theoretical asymptotic scaling exponent $2/9\approx 0.228$ can be attributed to the contribution to the heat transfer of the thermal boundary layer near the surface (see the discussion after equation (13) in Boeck \& Thess, 2001). Although this contribution is expected to vanish as $\Ma\to\infty$, the background method could yield a bound that agrees well with observations at least over a finite range of Marangoni numbers if the stability of the rolls were deduced rigorously from the governing equations. Given the lack of such information, however,~\eqref{e:SimpleCondition} must be satisfied for all wavenumbers up to the cut-off value $k_c=O(\Ma^{1/2})$. In particular, setting $k=1/\eps$ (which is no larger than $k_c$ by assumption) shows that we must choose $\eps \leq O(\Ma^{-1/2})$ and $\delta \leq O(\Ma^{-2/7})$, so the eventual bound on {\Nu} cannot grow more slowly than $O(\Ma^{2/7})$. The issue remains when we let $\gamma$ increase with {\Ma} to mimic the behaviour of the numerically optimal profiles (cf. figure~\ref{f:BLdataFull}{\it(d)}), because the apparent gain in~\eqref{e:finalbound} is exactly outbalanced by the need of testing wavenumbers up to $k_c =O(\gamma\Ma^{1/2})$. We therefore expect that to improve \HD's scaling using~\eqref{e:SimpleCondition} will require careful estimates of $a_k$, $b_k$ and $c_k$ at large wavenumbers, perhaps in conjunction with a more sophisticated choice of background field.

\subsection{Reaching the asymptotic regime: current challenges for conic optimisation}
\label{s:scalability}

As mentioned at the beginning of \S\ref{s:improvedbound}, the true asymptotic nature of our numerical bound remains uncertain due to the limited range of Marangoni numbers that could be studied. Clearly, this kind of uncertainty is inherent to any kind of numerical investigation irrespective of which computational tools are employed. Nonetheless, the challenges faced by conic programming in reaching the asymptotic regime deserve further discussion. 

The main limitation to extending the results presented in \S\ref{s:results} to larger values of {\Ma} is computational cost: proceeding from $\Ma=10^8$ to $\Ma = 10^9$ took more than 48 hours on our machine, and to achieve significant further progress would require computational resources beyond those available to the present investigation.
One difficulty is that at large Marangoni numbers checking whether a candidate background field satisfies the Fourier-transformed spectral constraints up to the cut-off wavenumber $k_c$ becomes a burden. For example, $k_c= 20\,073$ at $\Ma=10^9$, meaning that $20\,073$ eigenvalue decompositions must be computed after each iteration of the wavenumber-tracking procedure described in \S\ref{s:implementation}. 
The situation is worsened by the occurrence of bifurcations in critical wavenumbers, because more iterations are needed to correctly track all critical branches. Performance could be not improved by taking smaller steps in {\Ma}, because doing so slows progress towards higher Marangoni numbers. Increasing the parameter $r$ in~\eqref{e:critWavenoRange} also does not help much, because the cost of adding more LMIs to our conic programmes at each iteration offsets the reduction in number of iterations required to identify the critical wavenumbers. 

A possible solution to the critical wavenumber identification problem could be to apply the time-marching algorithm of~\citet{Wen2013,Wen2015a} to the optimality conditions for our conic programmes~\eqref{e:CP}--\eqref{e:CP2}. 
This method has been reported to locate the correct critical wavenumbers efficiently, although convergence to the optimal background field can be slow~\citep{Wen2015a}. Fast but less accurate solvers for conic programmes~\citep[such as {\sc SCS} by][]{ODonoghue2016} may also have similar benefits and drawbacks, with the additional advantage that finely tuned open-source implementations are readily available. Irrespectively of which method is utilised, once the critical wavenumbers have been identified the optimal solution can be computed using accurate conic programming packages such as SDPT3 (used in this work).

Our numerical method and the possible improvements discussed above can of course be applied beyond {\BM} convection. However, to study the asymptotic regime of more complex background method problems will require overcoming some additional obstacles.  Spectral constraints with multiple test functions, such as those encountered in shear flows~\citep{Plasting2003,Fantuzzi2016PRE} or finite-Prandtl-number convection~\citep{Doering1996,Otero2002}, yield conic programmes with larger LMIs. While current state-of-the-art algorithms for conic programming can handle many small LMIs very efficiently, the computational cost of a single LMI grows as a nonlinear function of its size. This problem is exacerbated for problems with two- and higher-dimensional background fields that cannot be Fourier-transformed in the horizontal directions, because after discretisation one obtains a single LMI instead of a set of smaller, independent LMIs corresponding to each wavevector.

On the other hand, the (current) unfavourable scalability of algorithms for conic programming can be mitigated by taking advantage of special properties of the particular background field problem at hand. For instance, the spectral constraint often presents symmetries that can be exploited to reduce the number of degrees of freedom needed to discretise the background field or the test functions (however, this is not the case for {\BM} convection). In addition, the very choice of discretisation method plays an important role because it directly impacts the sparsity of the eventual LMI. The piecewise-linear approximation method considered in this work is particularly attractive in this respect because it results in a {\it chordal} sparsity pattern, meaning that the nonzero entries of the LMI approximation of the spectral constraint can be represented by a chordal graph~\citep[a thorough discussion of these concepts is beyond the scope of this work, and we refer the interested reader to][Section 2]{Fukuda2000}.  The same is true when one uses multidimensional piecewise-polynomial representations in the spirit of finite-element methods. Chordal sparsity enables one to decompose a large LMI into multiple smaller ones, at the expense of introducing extra optimisation variables~\citep{Fukuda2000,Nakata2003,Kim2011}. This procedure can be automated,  for instance using the MATLAB toolbox SparseCoLO~\citep{Fujisawa2009}. As mentioned above, current algorithms for conic programming can handle multiple small LMIs much more efficiently than a single large one. Decomposition techniques based on chordal sparsity proved extremely effective in our study of {\BM} convection and we expect the same to be true for other background method problems.

Finally, the development of efficient algorithms for large-scale conic programmes and implementations that take advantage of modern parallel computer architectures is a very active area of research~\citep{Sun2014,Pakazad2015,Madani2015,ODonoghue2016,Zheng2016ifac,Zheng2016acc}. While it remains imperative to exploit all available symmetries and sparsity, advances at the algorithmic level promise to extend the ability of conic programming  to reach the asymptotic regime of background method problems more complex than the one considered in this work.

\section{Conclusion}
\label{s:conclusion}

This work studied the vertical heat transfer in {\BM} convection of a fluid layer with infinite Prandtl number by means of rigorous upper bounds on the Nusselt number. 
First, the background method analysis by \citet{Hagstrom2010} was extended to include balance parameters and formulate a new variational principle for the bound. Using this we proved that $\Nu\leq 0.803\times \Ma^{2/7}$, reducing the prefactor of the previous best bound by approximately 4.2\%, but we also showed that optimising the balance parameters does not affect the asymptotic scaling of the optimal bounds compared to \HD's original formulation. We then employed conic programming to optimise the bound on {\Nu} over all background fields, as well as over two smaller families constrained by either a monotonicity or a convexity constraint. 
The main result of our numerical investigation was the observation that the fully optimal bounds have the form $\Nu \leq O(\Ma^{2/7}(\ln\Ma)^{-1/2})$ for large Marangoni numbers. We also demonstrated that to achieve a logarithmic bound requires a background field with a non-monotonic boundary layer near the surface of the fluid. 
%
%

Whether the logarithmic scaling observed numerically can be proven analytically remains an open question, and is the subject of ongoing research.
The analysis presented in \S\ref{s:discussion} suggests a way forward by replacing the spectral constraint on the background field with the sufficient condition~\eqref{e:SimpleCondition}. Using~\eqref{e:SimpleCondition} is an attractive option because it is easier to check than the spectral constraint for a candidate background field, and the role of non-monotonicity is apparent. Moreover, the fact that enforcing~\eqref{e:SimpleCondition} at large wavenumbers seems to constrain the bound on {\Nu} is reminiscent of the bifurcations in critical wavenumbers observed in our numerical investigation (cf. figure~\ref{f:bifurcations}). In summary, condition~\eqref{e:SimpleCondition} seems to capture the essential features of the spectral constraint.

Should~\eqref{e:SimpleCondition} prove too strong, the analysis of the energy stability problem~\citep{Fantuzzi2017a} may be adapted to derive an inequality that exactly enforces each Fourier-transformed spectral constraint. The disadvantage is that such an inequality may not be analytically tractable except for very simple choices of the background field.
On the other hand, it may be possible to check this condition numerically and confirm that a candidate background field can indeed achieve a logarithmic bound, leaving ``only'' the task of constructing the correct estimates to prove so rigorously.
Alternatively, one may consider the Lagrangian dual of the variational problem obtained with background method. This amounts to constructing the temperature and velocity fields that maximise the heat transfer subject to the linearised momentum equation, the boundary conditions, and suitably averaged versions of the advection-diffusion equation for the temperature~\citep[for a detailed discussion of the duality between these two approaches in the context of Rayleigh--B\'enard convection, we refer the reader to][]{Plasting2005}. However, only the fields achieving the maximal heat transfer yield a fully rigorous bound, so the maximisation must be solved exactly. Moreover, compared to the Rayleigh--B\'enard problem the construction of a suitable  hierarchy of nested boundary layers  using Busse's ``multi-$\alpha$'' solution method~\citep[see for example][]{Busse1979} is complicated by the lack of vertical symmetry and the Neumann conditions at the surface of the fluid.


Irrespectively of how the variational problem for the upper bound on {\Nu} is analysed, however, it is evident from the present numerical investigation that the background method for the temperature field (or its dual formulation) cannot close the gap with the phenomenological prediction $\Nu=O(\Ma^{2/9})$ by~\cite{Boeck2001}. It is possible that Boeck \& Thess's assumption that steady convection rolls remain stable as $\Ma\to\infty$ is incorrect, making a scaling exponent of 2/9 unattainable with any bounding method.
To prove so rigorously requires a lower bound on {\Nu} that grows faster than $\Ma^{2/9}$, which can also not be achieved with the background method because the unstable conduction solution saturates the constant lower bound $\Nu\geq 1$.
Consequently, further numerical simulations in the high-{\Ma} seem essential to investigate the issue. The observation of steady convection rolls would provide further supporting evidence for Boeck \& Thess's phenomenological prediction. Determining the stability of the steady rolls is of interest also to reveal if the bifurcations in critical wavenumbers observed in our computations correspond to yet unobserved physical instabilities.

If Boeck \& Thess's phenomenological prediction is corroborated by further DNSs, to confirm it through rigorous bounds on {\Nu} will necessarily require bounding techniques beyond the background method. Unfortunately, the formulation of a wall-to-wall optimal transport problem in the spirit of~\citet{Hassanzadeh2014} and~\cite{Tobasco2017} does not appear suited to the study {\BM} convection at infinite-{\Pran}. In fact, the optimal transport approach treats the temperature as a passively advected and diffusing scalar, and one looks for the (generally time-dependent) incompressible velocity field that maximises the passive vertical transport of heat subject to a maximum power budget. However, in infinite-{\Pran} {\BM} convection the flow velocity is a linear function of the temperature field, which is effectively the only dynamical variable. This coupling is crucial in the background method analysis, so improving our bound on {\Nu} without taking it into account seems unlikely.

It would then be tempting to formulate the ``ultimate'' optimal wall-to-wall transport problem using the temperature as the decision variable, and let the flow velocity be specified as a function of it. However, this corresponds to searching for the exact solution of the equations of motion~\eqref{e:NSmomentum}--\eqref{e:continuity} with maximal heat transfer, so any significant progress does not appear possible. Difficulties remain when one drops the time dependence: maximising the heat transfer among the steady solutions is not much easier, and in any case the eventual bound would rely on the unproven assumption that unsteady flows cannot transport more heat than steady ones. Nonetheless, the construction of {\Ma}-dependent exact solutions remains of interest because knowledge of a (possibly unstable) flow with Nusselt number $Nu_{\rm ss}$ places a strict limit on what can be achieved by upper-bounding theory. In particular, any bounds that apply equally to all solutions of~\eqref{e:NSmomentum}--\eqref{e:continuity} cannot be better than $\Nu_{\rm ss}$.  Moreover, any flow with heat transfer $\Nu_{\rm ss} \gg O(\Ma^{2/9})$ would demonstrate that Boeck \& Thess's phenomenological scaling applies at most to a particular subset of all possible convective flows.

While improving the rigorous upper bound on {\Nu} using the ``ultimate'' wall-to-wall optimal transport approach described above appears challenging, it may be possible to consider successively weaker, tractable relaxations of it. The idea stems from the aforementioned realisation that the background method analysis is dual to the problem of maximising the heat transfer over all temperature (and associated velocity) fields that satisfy a set of constraints obtained by averaging the heat equation~\citep{Plasting2005}. The upper bound on {\Nu} may therefore be improved by including additional constraints implied by the heat equation, but not the heat equation itself. A simple way to do so is through a general bounding framework that encompasses the background method~\citep{Chernyshenko2014a,Chernyshenko2017}. The essence of this approach is to construct a functional $\mathcal{V}$ of the flow variables subject to a positivity condition akin to the spectral constraint in the background method. Each term in this functional can be interpreted as enforcing a particular constraint implied by the governing equations. Taking $\mathcal{V}$ to be the volume average of a quadratic polynomial of the flow variables gives the same bound as the background method~\citep{Chernyshenko2017}, but experience with finite-dimensional systems~\citep{Fantuzzi2016siads,Goluskin2016arxiv} indicates that considering more general functionals---for instance, volume averages of higher-than-quadratic polynomials of the flow variables---could yield significant improvements. Although the construction of suitable functionals may be beyond the reach of purely analytical work, progress can be assisted by computations that utilise conic programming techniques similar to those applied in this paper. Whether the numerical bounds can reach the asymptotic regime is of course highly dependent on the availability of efficient algorithmic tools for conic programming. Promising recent developments in this field~\citep[see for example][]{ODonoghue2016,Zheng2016ifac,Zheng2016acc}, however, give us hope that 
B\'enard--Marangoni convection and other turbulent hydrodynamic systems may  be studied successfully in the near future.

\appendix

\section{Minimisation of $\mathcal{Q}_0\{\hat{\theta}_0\}$}
\label{a:infQ0}

Let $\hat{\theta}_0(z)=v(z)$ to simplify the notation. It is not difficult to check using the calculus of variations that the infimum of $\mathcal{Q}_0$ over all test functions $v$ that satisfy $v(0)=0$ and $v'(1)=0$ is not attained unless $\beta=2$. This difficulty can be resolved by noticing that 
\begin{equation}
\label{e:NeumannDirichletEquiv}
\inf_{\substack{v(0)=0,\\v'(1)=0}} \mathcal{Q}_0\{v\} 
= \min_A \min_{\substack{v(0)=0,\\v(1)=A}} \mathcal{Q}_0\{v\}.
%
\end{equation}
In other words, we can replace the Neumann BC $v'(0)=0$ with the Dirichlet condition $v(1)=A$, solve the Dirichlet problem
\begin{equation}
\label{e:DirichletProblem}
\mathcal{Q}_0^\star(A) \defeq \min_{\substack{v(0)=0,\\v(1)=A}} \mathcal{Q}_0\{v\},
\end{equation}
and minimise $\mathcal{Q}_k^\star(A)$ over $A$. 
Equation~\eqref{e:NeumannDirichletEquiv} is justified because for each value $A$, the minimum of the Dirichlet problem can be approximated with arbitrary accuracy by a function that satisfies $v'(1)=0$; for example, if $v^\star$ is the minimiser of the Dirichlet problem~\eqref{e:DirichletProblem} for a given $A$, take
\begin{equation}
v(z) = \begin{cases} 
v^\star(z), &0\leq z \leq 1-\delta,\\
v^\star(1-\delta), &1-\delta\leq z \leq 1
\end{cases}
\end{equation}
for $\delta>0$ sufficiently small.
A rigorous proof is omitted for brevity, but a similar argument can be found in a previous work by the authors~\citep[appendix C]{Fantuzzi2017a}.

The minimiser of the Dirichlet problem~\eqref{e:DirichletProblem} satisfies the Euler--Lagrange equation
\begin{equation}
-2\,v'' - \frac{\alpha-2}{\alpha-1}\,\tau'' = 0
\end{equation}
subject to the BCs $v(0)=0$ and $v(1)=A$, and is given by
\begin{equation}
v^\star(z) = \frac{\alpha-2}{2(\alpha-1)}\left[ \tau(1)\,z - \tau(z) \right] + A\,z.
\end{equation}
The corresponding minimum is
\begin{equation}
\mathcal{Q}_0^\star(A) = 
A^2 + \frac{ 
		(\alpha-2)\,\tau(1)
		+\alpha-\beta}{\alpha-1}\,A
	+\frac{(\alpha-2)^2\left[ \abs{\tau(1)}^2-\normtwo{\tau'}^2 \right]}{4(\alpha-1)^2}.
\end{equation}
An expression for the minimum over $A$ is readily found, and it can be rearranged in the form~\eqref{e:infQ0} after noticing that
$\tau(1) = \int_0^1 \tau'(z) \dz$ by virtue of~\eqref{e:bfBC}.

\section{An improved bound on {\Nu}}
\label{a:HDproof}

Consider a piecewise-linear scaled background field of the form
\begin{equation}
\label{e:PLrho}
\rho(z) = \begin{cases}
-R\,z &0\leq z \leq \delta,\\
-R\,\delta, & \delta\leq z \leq 1.
\end{cases}
\end{equation}
The boundary layer slope $R>0$ and thickness $\delta>0$ should be chosen to satisfy the spectral constraint~\eqref{e:SC} whilst optimising the bound on the Nusselt number,
\begin{equation}
\label{e:PLbound}
\frac{1}{\Nu} \geq 
\frac{1 - \normtwo{{\rho}'+1} - \rho(1)}{2} 
= \frac{1 - \sqrt{1+R\,(R-2)\,\delta} +R\,\delta}{2}.
\end{equation}

Recall from \S\ref{s:BackgroundMethod} that the spectral constraint is equivalent to the quadratic form $\mathcal{Q}_k\{\hat{\theta}_k\}$ in~\eqref{e:QkDef} being positive semidefinite for all wavenumbers $k\geq 1$, and recall that we have changed variables such that $\alpha/(\alpha-1)\tau'(z) = \rho'(z)$. Although the test function $\hat{\theta}_k$ is complex-valued, the contributions of its real and imaginary parts to $\mathcal{Q}_k\{\hat{\theta}_k\}$ are identical and independent, so it suffices to consider real-valued test functions. We conclude that  $R$ and $\delta$ must be chosen such that, for all $k\geq 1$,
\begin{equation}
\label{e:PLSC}
\mathcal{Q}_k\{v\} = \normtwo{v'}^2 + k^2\,\normtwo{v}^2 - \Ma\,R\,v(1) \int_0^\delta f_k(z)\,v(z)\dz \geq 0
\end{equation}
for all real-valued functions $v(z)$ that satisfy the BCs $v(0)=0$ and $v'(1)=0$. 

Using the BC $v(0)=0$ and the Cauchy--Schwarz inequality, we can bound
\begin{equation}
\abs{v(1)} = \abs{\int_0^1 v'(z) \dz} \leq \normtwo{v'}^2.
\end{equation}
Moreover, since $\abs{f_k(z)} = -f_k(z)\leq c\,z^2$ for $c\approx 0.943$~\citep{Hagstrom2010}, 
\begin{align}
\abs{\Ma\,R\,v(1)\int_0^\delta f_k(z)\,v(z)\dz}  
&\leq \Ma\,R\,c\,\abs{\int_0^\delta \int_0^z z^2\,v'(\xi)\dxi \dz}\,\normtwo{v'}
\nonumber \\
&= \Ma\,R\,c\,\abs{\int_0^\delta \int_\xi^\delta z^2\,v'(\xi)\dxi \dz}\,\normtwo{v'}
\nonumber \\
&= \frac{\Ma\,R\,c}{3} \abs{\int_0^\delta \left(\delta^3-\xi^3\right)\,v'(\xi)\dxi \dz}\,\normtwo{v'}
\nonumber \\
&\leq \frac{\Ma\,R\,c}{3}\sqrt{\int_0^\delta \left(\delta^3-\xi^3\right)^2 \dxi} \,\normtwo{v'}^2
\nonumber \\
&= \frac{\Ma\,R\,c\,\delta^{7/2}}{\sqrt{14}} \,\normtwo{v'}^2.
\end{align}
Inequality~\eqref{e:PLSC} therefore holds if
\begin{equation}
\delta = \left( \frac{\Ma\,R\,c}{\sqrt{14}} \right)^{-2/7}.
\end{equation}
With this choice of $\delta$, the asymptotic behaviour of the bound~\eqref{e:PLbound} as the Marangoni number tends to infinity is
\begin{equation}
\frac{1}{\Nu} \geq
\left(\frac{\sqrt{14}}{c}\right)^{2/7}\frac{R\,(4-R)}{4\,R^{2/7}} \Ma^{-2/7},
\end{equation}
and choosing $R=5/3$ to maximize the prefactor we arrive at
\begin{equation}
\Nu \leq \frac{36}{35}\,\left(\frac{5\,c}{3\,\sqrt{14}}\right)^{2/7} \Ma^{2/7} \approx 0.803\, \Ma^{2/7}
\quad
\text{as } \Ma\to\infty.
\end{equation}

\section{Computation of the cut-off wavenumber $k_{\rm c}$}
\label{ss:kCritBound}

\begin{figure}
\centerline{\includegraphics{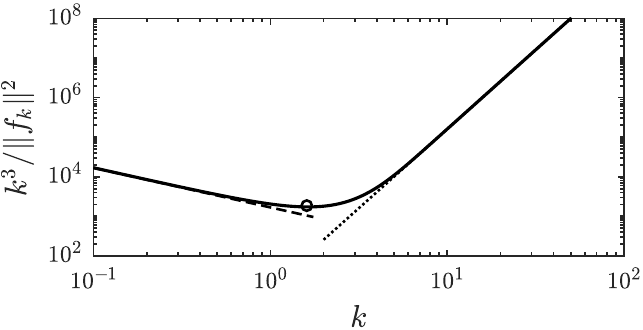}}
\caption{The quantity $k^3/\normtwo{f_k}^2$ (solid line) with its large- and small-wavenumber asymptotes (dotted and dashed lines, respectively). A circle marks the minimum at $k\approx1.633$.}
\label{f:kBound}
\end{figure}

Since any test function $v\in\Gamma$ vanishes at $z=0$, integration by parts shows that for any constant $\gamma \geq 0$
\begin{equation}
\label{e:IP1}
\gamma\abs{v(1)}^2 
-2\,\gamma\int_0^1\!v\,v'\dz = 0.
\end{equation}
Adding this to the quadratic form $\mathcal{Q}_k\{v\}$ in~\eqref{e:FTSCphi} and using the Cauchy--Schwarz inequality to estimate the sign-indefinite terms yields
\begin{multline}
\label{e:lbQk}
\mathcal{Q}_k\{v\}\! 
\geq 
\normtwo{v'}^2 + k^2\normtwo{v}^2 + \gamma\abs{v(1)}^2 
- 2\,\gamma\normtwo{v'}\normtwo{v}
\\
- \Ma\normtwo{(\phi-1)\,f_k}\abs{v(1)}\normtwo{v}.
\end{multline}
Consequently, $\mathcal{Q}_k\{v\}\geq 0$ if there exists a scalar $\omega$ such that
\begin{subequations}
\begin{align}
\label{e:kBound1} 
\normtwo{v'}^2 - 2\,\gamma\normtwo{v'}\normtwo{v} + \omega k^2\normtwo{v}^2 &\geq 0,
\\ 
\left(1-\omega\right) k^2 \normtwo{v}^2
- \Ma\normtwo{(\phi-1)f_k}\abs{v(1)}\normtwo{v}
+ \gamma\abs{v(1)}^2  &\geq 0.
\end{align}
\end{subequations}
Recalling that a quadratic form $a x^2 + b x y + c y^2$ is non-negative for all $x$ and $y$ if $b^2\leq 4ac$, choosing $\omega>0$ and $\gamma = \sqrt{\omega}k$ to complete the square in~\eqref{e:kBound1} implies that $\mathcal{Q}_k\{v\}\geq 0$ if
\begin{equation}
\Ma^2 \normtwo{(\phi-1)\,f_k}^2 
\leq 4\,\left(1-\omega\right)\sqrt{\omega}\,k^3.
\end{equation}
After setting $\omega=1/3$ to maximise the right-hand side, estimating
\begin{equation}
\normtwo{(\phi-1)\,f_k}\leq \norm{\phi-1}_\infty \normtwo{f_k},
\end{equation} 
and rearranging, we arrive at
\begin{equation}
\label{e:kBound3}
\frac{k^3}{\normtwo{f_k}^2} \geq \frac{3\sqrt{3}}{8}\,\Ma^2 \norm{\phi-1}_\infty^2.
\end{equation}
As illustrated in figure~\ref{f:kBound}, the quantity $k^3/\normtwo{f_k}^2$ has a minimum 
at $k=k_{\rm crit}\approx1.633$, grows asymptotically to $1680\,k^{-1}$ as $k\to 0$, and quickly asymptotes $16\,k^4$ for $k>k_{\rm crit}$. In fact $k^3/\normtwo{f_k}^2\geq 16 k^4$ so~\eqref{e:kBound3}---and hence the spectral constraint---holds for all wavenumbers larger than the critical value
\begin{equation}
\begin{aligned}
k_c \defeq 
\left\lfloor 
\left(\frac{3\sqrt{3}}{128}\right)^{1/4} \Ma^{1/2} \norm{\phi-1}_\infty^{1/2} 
\right\rfloor.
\end{aligned}
\end{equation}
%

\bibliographystyle{./jfm}
\bibliography{refs}

\begin{thebibliography}{61}
\expandafter\ifx\csname natexlab\endcsname\relax\def\natexlab#1{#1}\fi
\def\au#1{#1} \def\ed#1{#1} \def\yr#1{#1}\def\at#1{#1}\def\jt#1{\textit{#1}}
  \def\bt#1{#1}\def\bvol#1{\textbf{#1}} \def\vol#1{#1} \def\pg#1{#1}
  \def\publ#1{#1}\def\arxiv#1{#1}\def\org#1{#1}\def\st#1{\textit{#1}}

\bibitem[Boeck(2005)]{Boeck2005}
{\sc \au{Boeck, T.}} \yr{2005}  \at{{B\'enard--Marangoni convection at large
  Marangoni numbers: Results of numerical simulations}}.  \jt{Adv. Sp. Res.}
  \bvol{36}~(1),  \pg{4--10}.

\bibitem[Boeck \& Thess(1998)]{Boeck1998}
{\sc \au{Boeck, T.} \& \au{Thess, A.}} \yr{1998}  \at{{Turbulent
  B\'enard--Marangoni convection: Results of two-dimensional simulations}}.
  \jt{Phys. Rev. Lett.}  \bvol{80}~(6),  \pg{1216--1219}.

\bibitem[Boeck \& Thess(2001)]{Boeck2001}
{\sc \au{Boeck, T.} \& \au{Thess, A.}} \yr{2001}  \at{{Power-law scaling in
  B\'{e}nard--Marangoni convection at large Prandtl numbers}}.  \jt{Phys. Rev.
  E}  \bvol{64}~(2),  \pg{027303}.

\bibitem[Boyd {\em et~al.\/}(1994)Boyd, {El Ghaoui}, Feron \&
  Balakrishnan]{Boyd1994}
{\sc \au{Boyd, S.}, \au{{El Ghaoui}, L.}, \au{Feron, E.} \& \au{Balakrishnan,
  V.}} \yr{1994} {\em {Linear matrix inequalities in system and control
  theory}\/}.  \publ{SIAM}.

\bibitem[Boyd \& Vandenberghe(2004)]{Boyd2004}
{\sc \au{Boyd, S.} \& \au{Vandenberghe, L.}} \yr{2004} {\em {Convex
  optimization}\/}.  \publ{Cambridge University Press}.

\bibitem[de~Bruyn {\em et~al.\/}(1996)de~Bruyn, Bodenschatz, Morris, Trainoff,
  Hu, Cannell \& Ahlers]{DeBruyn1996}
{\sc \au{de~Bruyn, J.~R.}, \au{Bodenschatz, E.}, \au{Morris, S.~W.},
  \au{Trainoff, S.~P.}, \au{Hu, Y.}, \au{Cannell, D.~S.} \& \au{Ahlers, G.}}
  \yr{1996}  \at{{Apparatus for the study of Rayleigh--B{\'{e}}nard convection
  in gases under pressure}}.  \jt{Rev. Sci. Instrum.}  \bvol{67}~(6),
  \pg{2043--2067}.

\bibitem[Busse(1979)]{Busse1979}
{\sc \au{Busse, F.~H.}} \yr{1979}  \at{The optimum theory of turbulence}.
  \jt{Adv. Appl. Mech.}  \bvol{18},  \pg{77--121}.

\bibitem[Chernyshenko(2017)]{Chernyshenko2017}
{\sc \au{Chernyshenko, S.~I.}} \yr{2017} {Relationship between the methods of
  bounding time averages},  \arxiv{arXiv:1704.02475v2 [physics.flu-dyn]}.

\bibitem[Chernyshenko {\em et~al.\/}(2014)Chernyshenko, Goulart, Huang \&
  Papachristodoulou]{Chernyshenko2014a}
{\sc \au{Chernyshenko, S.~I.}, \au{Goulart, P.~J.}, \au{Huang, D.} \&
  \au{Papachristodoulou, A.}} \yr{2014}  \at{{Polynomial sum of squares in
  fluid dynamics: a review with a look ahead}}.  \jt{Philos. Trans. R. Soc. A}
  \bvol{372}~(2020),  \pg{20130350}.

\bibitem[Constantin \& Doering(1995{\natexlab{{\em a\/}}})]{Constantin1995}
{\sc \au{Constantin, P.} \& \au{Doering, C.~R.}} \yr{1995{\natexlab{{\em
  a\/}}}}  \at{{Variational bounds in dissipative systems}}.  \jt{Phys. D}
  \bvol{82}~(3),  \pg{221--228}.

\bibitem[Constantin \& Doering(1995{\natexlab{{\em b\/}}})]{Constantin1995a}
{\sc \au{Constantin, P.} \& \au{Doering, C.~R.}} \yr{1995{\natexlab{{\em
  b\/}}}}  \at{{Variational bounds on energy dissipation in incompressible
  flows. II. Channel flow}}.  \jt{Phys. Rev. E}  \bvol{51}~(4),
  \pg{3192--3198}.

\bibitem[DebRoy \& David(1995)]{Debroy1995}
{\sc \au{DebRoy, T.} \& \au{David, S.~A.}} \yr{1995}  \at{{Physical processes
  in fusion welding}}.  \jt{Rev. Mod. Phys.}  \bvol{67}~(1),  \pg{85--112}.

\bibitem[Doering \& Constantin(1992)]{Doering1992}
{\sc \au{Doering, C.~R.} \& \au{Constantin, P.}} \yr{1992}  \at{{Energy
  dissipation in shear driven turbulence}}.  \jt{Phys. Rev. Lett.}
  \bvol{69}~(11),  \pg{1648--1651}.

\bibitem[Doering \& Constantin(1994)]{Doering1994}
{\sc \au{Doering, C.~R.} \& \au{Constantin, P.}} \yr{1994}  \at{{Variational
  bounds on energy dissipation in incompressible flows: Shear flow}}.
  \jt{Phys. Rev. E}  \bvol{49}~(5),  \pg{4087--4099}.

\bibitem[Doering \& Constantin(1996)]{Doering1996}
{\sc \au{Doering, C.~R.} \& \au{Constantin, P.}} \yr{1996}  \at{{Variational
  bounds on energy dissipation in incompressible flows. III. Convection}}.
  \jt{Phys. Rev. E}  \bvol{53}~(6),  \pg{5957--5981}.

\bibitem[Doering {\em et~al.\/}(2006)Doering, Otto \& Reznikoff]{Doering2006}
{\sc \au{Doering, C.~R.}, \au{Otto, F.} \& \au{Reznikoff, M.~G.}} \yr{2006}
  \at{{Bounds on vertical heat transport for infinite-Prandtl-number
  Rayleigh--B\'enard convection}}.  \jt{J. Fluid Mech.}  \bvol{560},
  \pg{229--241}.

\bibitem[Eckert \& Thess(2006)]{Eckert2006}
{\sc \au{Eckert, K.} \& \au{Thess, A.}} \yr{2006}  \at{{Secondary instabilities
  in surface-tension-driven B{\'{e}}nard--Marangoni convection}}.  \bt{In {\em
  {Dynamics of spatio-temporal cellular structures}\/} (ed. \ed{I.~Mutabazi,
  J.~E. Wesfreid \& E.~Guyon})},  \st{{Springer tracts in modern physics}},
  \vol{vol. 207}, chap.~9,  \pg{pp. 163--176}.  \publ{Springer New York}.

\bibitem[Fantuzzi {\em et~al.\/}(2016)Fantuzzi, Goluskin, Huang \&
  Chernyshenko]{Fantuzzi2016siads}
{\sc \au{Fantuzzi, G.}, \au{Goluskin, D.}, \au{Huang, D.} \& \au{Chernyshenko,
  S.~I.}} \yr{2016}  \at{{Bounds for deterministic and stochastic dynamical
  systems using sum-of-squares optimization}}.  \jt{SIAM J. Appl. Dyn. Syst.}
  \bvol{15}~(4),  \pg{1962--1988}.

\bibitem[Fantuzzi \& Wynn(2015)]{Fantuzzi2015}
{\sc \au{Fantuzzi, G.} \& \au{Wynn, A.}} \yr{2015}  \at{{Construction of an
  optimal background profile for the Kuramoto--Sivashinsky equation using
  semidefinite programming}}.  \jt{Phys. Lett. A}  \bvol{379}~(1-2),
  \pg{23--32}.

\bibitem[Fantuzzi \& Wynn(2016{\natexlab{{\em a\/}}})]{Fantuzzi2016PRE}
{\sc \au{Fantuzzi, G.} \& \au{Wynn, A.}} \yr{2016{\natexlab{{\em a\/}}}}
  \at{{Optimal bounds with semidefinite programming: An application to stress
  driven shear flows}}.  \jt{Phys. Rev. E}  \bvol{93}~(4),  \pg{043308}.

\bibitem[Fantuzzi \& Wynn(2016{\natexlab{{\em b\/}}})]{Fantuzzi2016CDC}
{\sc \au{Fantuzzi, G.} \& \au{Wynn, A.}} \yr{2016{\natexlab{{\em b\/}}}}
  {Semidefinite relaxation of a class of quadratic integral inequalities}.
  \bt{In {\em Proc. 55th IEEE Annu. Conf. Decis. Control\/}},  \pg{pp.
  6192--6197}. Las Vegas, USA.

\bibitem[Fantuzzi \& Wynn(2017)]{Fantuzzi2017a}
{\sc \au{Fantuzzi, G.} \& \au{Wynn, A.}} \yr{2017}  \at{{Exact energy stability
  of B{\'{e}}nard--Marangoni convection at infinite Prandtl number}}.  \jt{J.
  Fluid Mech.}  \bvol{822},  \pg{R1}.

\bibitem[Fantuzzi {\em et~al.\/}(2017)Fantuzzi, Wynn, Goulart \&
  Papachristodoulou]{Fantuzzi2017TAC}
{\sc \au{Fantuzzi, G.}, \au{Wynn, A.}, \au{Goulart, P.~J.} \&
  \au{Papachristodoulou, A.}} \yr{2017}  \at{{Optimization with affine
  homogeneous quadratic integral inequality constraints}}.  \jt{Trans. Autom.
  Control} (in press, early access copy available from
  \url{https://doi.org/10.1109/TAC.2017.2703927}).

\bibitem[Fujisawa {\em et~al.\/}(2009)Fujisawa, Kim, Kojima, Okamoto \&
  Yamashita]{Fujisawa2009}
{\sc \au{Fujisawa, K.}, \au{Kim, S.}, \au{Kojima, M.}, \au{Okamoto, Y.} \&
  \au{Yamashita, M.}} \yr{2009}  \bt{{User's manual for SparseCoLO: conversion
  methods for SPARSE COnic-form Linear Optimization problems}}. {\em Tech.
  Rep.\/} B-453.  \org{Dept. of Mathematical and Computing Sciences, Tokyo
  Institute of Technology}, Tokyo, Japan.

\bibitem[Fukuda {\em et~al.\/}(2000)Fukuda, Kojima, Murota \&
  Nakata]{Fukuda2000}
{\sc \au{Fukuda, M.}, \au{Kojima, M.}, \au{Murota, K.} \& \au{Nakata, K.}}
  \yr{2000}  \at{{Exploiting sparsity in semidefinite programming via matrix
  completion I: General framework}}.  \jt{SIAM J. Optim.}  \bvol{11}~(3),
  \pg{647--674}.

\bibitem[Goluskin(2016)]{Goluskin2016arxiv}
{\sc \au{Goluskin, D.}} \yr{2016} {Bounding averages rigorously using
  semidefinite programming: mean moments of the Lorenz system},
  \arxiv{arXiv:1610.05335v1 [math.DS]}.

\bibitem[Goluskin \& Doering(2016)]{Goluskin2016d}
{\sc \au{Goluskin, D.} \& \au{Doering, C.~R.}} \yr{2016}  \at{{Bounds for
  convection between rough boundaries}}.  \jt{J. Fluid Mech.}  \bvol{804},
  \pg{370--386}.

\bibitem[Hagstrom \& Doering(2010)]{Hagstrom2010}
{\sc \au{Hagstrom, G.} \& \au{Doering, C.~R.}} \yr{2010}  \at{{Bounds on heat
  transport in B{\'{e}}nard--Marangoni convection}}.  \jt{Phys. Rev. E}
  \bvol{81}~(4),  \pg{047301}.

\bibitem[Hassanzadeh {\em et~al.\/}(2014)Hassanzadeh, Chini \&
  Doering]{Hassanzadeh2014}
{\sc \au{Hassanzadeh, P.}, \au{Chini, G.~P.} \& \au{Doering, C.~R.}} \yr{2014}
  \at{{Wall to wall optimal transport}}.  \jt{J. Fluid Mech.}  \bvol{751},
  \pg{627--662}.

\bibitem[Kim {\em et~al.\/}(2011)Kim, Kojima, Mevissen \& Yamashita]{Kim2011}
{\sc \au{Kim, S.}, \au{Kojima, M.}, \au{Mevissen, M.} \& \au{Yamashita, M.}}
  \yr{2011}  \at{{Exploiting sparsity in linear and nonlinear matrix
  inequalities via positive semidefinite matrix completion}}.  \jt{Math.
  Program. Ser. B}  \bvol{129}~(1),  \pg{33--68}.

\bibitem[Kumar \& Roy(2009)]{Kumar2009}
{\sc \au{Kumar, A.} \& \au{Roy, S.}} \yr{2009}  \at{{Effect of
  three-dimensional melt pool convection on process characteristics during
  laser cladding}}.  \jt{Comput. Mater. Sci.}  \bvol{46}~(2),  \pg{495--506}.

\bibitem[Lappa(2010)]{Lappa2010}
{\sc \au{Lappa, M.}} \yr{2010} {\em {Thermal convection: patterns, evolution
  and stability}\/}.  \publ{John Wiley {\&} Sons, Ltd}.

\bibitem[L\"ofberg(2004)]{Lofberg2004}
{\sc \au{L\"ofberg, J.}} \yr{2004} {YALMIP: A toolbox for modeling and
  optimization in MATLAB}.  \bt{In {\em IEEE Int. Symp. Comput. Aided Control
  Syst. Des.\/}},  \pg{pp. 284--289}. Taipei, TW.

\bibitem[Madani {\em et~al.\/}(2015)Madani, Kalbat \& Lavaei]{Madani2015}
{\sc \au{Madani, R.}, \au{Kalbat, A.} \& \au{Lavaei, J.}} \yr{2015} {ADMM for
  sparse semidefinite programming with applications to optimal power flow
  problem}.  \bt{In {\em Proc. 54th IEEE Conf. Decis. Control\/}},  \pg{pp.
  5932--5939}.  \publ{Osaka, Japan: IEEE}.

\bibitem[Nakata {\em et~al.\/}(2003)Nakata, Fujisawa, Fukuda, Kojima \&
  Murota]{Nakata2003}
{\sc \au{Nakata, K.}, \au{Fujisawa, K.}, \au{Fukuda, M.}, \au{Kojima, M.} \&
  \au{Murota, K.}} \yr{2003}  \at{{Exploiting sparsity in semidefinite
  programming via matrix completion II: Implementation and numerical results}}.
   \jt{Math. Program. Ser. B}  \bvol{95}~(2),  \pg{303--327}.

\bibitem[Nicodemus {\em et~al.\/}(1997)Nicodemus, Grossmann \&
  Holthaus]{Nicodemus1998c}
{\sc \au{Nicodemus, R.}, \au{Grossmann, S.} \& \au{Holthaus, M.}} \yr{1997}
  \at{{Improved variational principle for bounds on energy dissipation in
  turbulent shear flow}}.  \jt{Phys. D}  \bvol{101}~(1--2),  \pg{178--190}.

\bibitem[O'Donoghue {\em et~al.\/}(2016)O'Donoghue, Chu, Parikh \&
  Boyd]{ODonoghue2016}
{\sc \au{O'Donoghue, B.}, \au{Chu, E.}, \au{Parikh, N.} \& \au{Boyd, S.}}
  \yr{2016}  \at{{Conic optimization via operator splitting and homogeneous
  self-dual embedding}}.  \jt{J. Optim. Theory Appl.}  \bvol{169}~(3),
  \pg{1042--1068}.

\bibitem[Otero(2002)]{Otero2002}
{\sc \au{Otero, J.}} \yr{2002}  \at{{Bounds for the heat transport in turbulent
  convection}}. PhD thesis, University of Michigan.

\bibitem[Otto \& Seis(2011)]{Otto2011}
{\sc \au{Otto, F.} \& \au{Seis, C.}} \yr{2011}  \at{{Rayleigh--B\'enard
  convection: Improved bounds on the Nusselt number}}.  \jt{J. Math. Phys.}
  \bvol{52}~(8),  \pg{083702}.

\bibitem[Pakazad {\em et~al.\/}(2015)Pakazad, Hansson, Andersen \&
  Rantzer]{Pakazad2015}
{\sc \au{Pakazad, S.~K.}, \au{Hansson, A.}, \au{Andersen, M.~S.} \&
  \au{Rantzer, A.}} \yr{2015}  \at{{Distributed semidefinite programming with
  application to large-scale system analysis}}.  \jt{IEEE Trans. Automat.
  Contr.} (in press, early access copy available from
  \url{https://doi.org/10.1109/TAC.2017.2739644}).

\bibitem[Patberg {\em et~al.\/}(1983)Patberg, Koers, Steenge \&
  Drinkenburg]{Patberg1983}
{\sc \au{Patberg, W.~B.}, \au{Koers, A.}, \au{Steenge, W. D.~E.} \&
  \au{Drinkenburg, A. A.~H.}} \yr{1983}  \at{{Effectiveness of mass transfer in
  a packed distillation column in relation to surface tension gradients}}.
  \jt{Chem. Eng. Sci.}  \bvol{38}~(6),  \pg{917--923}.

\bibitem[Pearson(1958)]{Pearson1958}
{\sc \au{Pearson, J. R.~A.}} \yr{1958}  \at{{On convection cells induced by
  surface tension}}.  \jt{J. Fluid Mech.}  \bvol{4}~(5),  \pg{489--500}.

\bibitem[Plasting(2004)]{Plasting2004}
{\sc \au{Plasting, S.~C.}} \yr{2004}  \at{{Turbulence has its limits: {\it a
  priori} estimates of transport properties in turbulent fluid flows}}. {PhD}
  thesis, University of Bristol.

\bibitem[Plasting \& Ierley(2005)]{Plasting2005}
{\sc \au{Plasting, S.~C.} \& \au{Ierley, G.~R.}} \yr{2005}
  \at{{Infinite-Prandtl-number convection. Part 1. Conservative bounds}}.
  \jt{J. Fluid Mech.}  \bvol{542}~(2005),  \pg{343--363}.

\bibitem[Plasting \& Kerswell(2003)]{Plasting2003}
{\sc \au{Plasting, S.~C.} \& \au{Kerswell, R.~R.}} \yr{2003}  \at{{Improved
  upper bound on the energy dissipation rate in plane Couette flow: the full
  solution to Busse's problem and the Constantin--Doering--Hopf problem with
  one-dimensional background field}}.  \jt{J. Fluid Mech.}  \bvol{477},
  \pg{363--379}.

\bibitem[Pumir \& Blumenfeld(1996)]{Pumir1996}
{\sc \au{Pumir, A.} \& \au{Blumenfeld, L.}} \yr{1996}  \at{{Heat transport in a
  liquid layer locally heated on its free surface}}.  \jt{Phys. Rev. E}
  \bvol{54}~(5),  \pg{R4528--R4531}.

\bibitem[Schatz \& Neitzel(2001)]{Schatz2001}
{\sc \au{Schatz, M.~F.} \& \au{Neitzel, G.~P.}} \yr{2001}  \at{{Experiments on
  thermocapillary instabilities}}.  \jt{Annu. Rev. Fluid Mech.}  \bvol{33},
  \pg{93--127}.

\bibitem[Sun {\em et~al.\/}(2014)Sun, Andersen \& Vandenberghe]{Sun2014}
{\sc \au{Sun, Y.}, \au{Andersen, M.~S.} \& \au{Vandenberghe, L.}} \yr{2014}
  \at{{Decomposition in conic optimization with partially separable
  structure}}.  \jt{SIAM J. Optim.}  \bvol{24}~(2),  \pg{873--897}.

\bibitem[Tobasco \& Doering(2017)]{Tobasco2017}
{\sc \au{Tobasco, I.} \& \au{Doering, C.~R.}} \yr{2017}  \at{{Optimal
  wall-to-wall transport by incompressible flows}}.  \jt{Phys. Rev. Lett.}
  \bvol{118}~(26),  \pg{264502}.

\bibitem[Toh {\em et~al.\/}(1999)Toh, Todd \&
  T{\"{u}}t{\"{u}}nc{\"{u}}]{Toh1999}
{\sc \au{Toh, K.~C.}, \au{Todd, M.~J.} \& \au{T{\"{u}}t{\"{u}}nc{\"{u}},
  R.~H.}} \yr{1999}  \at{{SDPT3 --- A MATLAB software package for semidefinite
  programming, version 1.3}}.  \jt{Optim. Methods Softw.}  \bvol{11}~(1--4),
  \pg{545--581}.

\bibitem[T{\"{u}}t{\"{u}}nc{\"{u}} {\em
  et~al.\/}(2003)T{\"{u}}t{\"{u}}nc{\"{u}}, Toh \& Todd]{Tutuncu2003}
{\sc \au{T{\"{u}}t{\"{u}}nc{\"{u}}, R.~H.}, \au{Toh, K.~C.} \& \au{Todd,
  M.~J.}} \yr{2003}  \at{{Solving semidefinite-quadratic-linear programs using
  SDPT3}}.  \jt{Math. Program. Ser. B}  \bvol{95}~(2),  \pg{189--217}.

\bibitem[Vandenberghe \& Boyd(1996)]{Vandenberghe1996}
{\sc \au{Vandenberghe, L.} \& \au{Boyd, S.}} \yr{1996}  \at{{Semidefinite
  programming}}.  \jt{SIAM Rev.}  \bvol{38}~(1),  \pg{49--95}.

\bibitem[Wen {\em et~al.\/}(2013)Wen, Chini, Dianati \& Doering]{Wen2013}
{\sc \au{Wen, B.}, \au{Chini, G.~P.}, \au{Dianati, N.} \& \au{Doering, C.~R.}}
  \yr{2013}  \at{{Computational approaches to aspect-ratio-dependent upper
  bounds and heat flux in porous medium convection}}.  \jt{Phys. Lett. A}
  \bvol{377}~(41),  \pg{2931--2938}.

\bibitem[Wen {\em et~al.\/}(2015)Wen, Chini, Kerswell \& Doering]{Wen2015a}
{\sc \au{Wen, B.}, \au{Chini, G.~P.}, \au{Kerswell, R.~R.} \& \au{Doering,
  C.~R.}} \yr{2015}  \at{{Time-stepping approach for solving upper-bound
  problems: Application to two-dimensional Rayleigh--B\'enard convection}}.
  \jt{Phys. Rev. E}  \bvol{92}~(4),  \pg{043012}.

\bibitem[Whitehead \& Doering(2011)]{Whitehead2011}
{\sc \au{Whitehead, J.~P.} \& \au{Doering, C.~R.}} \yr{2011}  \at{{Ultimate
  state of two-dimensional Rayleigh--B{\'{e}}nard convection between free-slip
  fixed-temperature boundaries}}.  \jt{Phys. Rev. Lett.}  \bvol{106}~(24),
  \pg{244501}.

\bibitem[Whitehead \& Doering(2012)]{Whitehead2012}
{\sc \au{Whitehead, J.~P.} \& \au{Doering, C.~R.}} \yr{2012}  \at{{Rigid bounds
  on heat transport by a fluid between slippery boundaries}}.  \jt{J. Fluid
  Mech.}  \bvol{707},  \pg{241--259}.

\bibitem[Wittenberg \& Gao(2010)]{Wittenberg2010a}
{\sc \au{Wittenberg, R.~W.} \& \au{Gao, J.}} \yr{2010}  \at{{Conservative
  bounds on Rayleigh--B\'{e}nard convection with mixed thermal boundary
  conditions}}.  \jt{Eur. Phys. J. B}  \bvol{76}~(4),  \pg{565--580}.

\bibitem[Yiantsios {\em et~al.\/}(2015)Yiantsios, Serpetsi, Doumenc \&
  Guerrier]{Yiantsios2015}
{\sc \au{Yiantsios, S.~G.}, \au{Serpetsi, S.~K.}, \au{Doumenc, F.} \&
  \au{Guerrier, B.}} \yr{2015}  \at{{Surface deformation and film corrugation
  during drying of polymer solutions induced by Marangoni phenomena}}.
  \jt{Int. J. Heat Mass Transf.}  \bvol{89},  \pg{1083--1094}.

\bibitem[Zheng {\em et~al.\/}(2017{\natexlab{{\em a\/}}})Zheng, Fantuzzi,
  Papachristodoulou, Goulart \& Wynn]{Zheng2016ifac}
{\sc \au{Zheng, Y.}, \au{Fantuzzi, G.}, \au{Papachristodoulou, A.},
  \au{Goulart, P.~J.} \& \au{Wynn, A.}} \yr{2017{\natexlab{{\em a\/}}}} {Fast
  ADMM for homogeneous self-dual embedding of sparse SDPs}.  \bt{In {\em Proc.
  20th World Congr. Int. Fed. Autom. Control\/}},  \pg{pp. 8741--8746}.
  Toulouse, France, to appear (pre-print available from
  \url{https://arxiv.org/abs/1611.01828}).

\bibitem[Zheng {\em et~al.\/}(2017{\natexlab{{\em b\/}}})Zheng, Fantuzzi,
  Papachristodoulou, Goulart \& Wynn]{Zheng2016acc}
{\sc \au{Zheng, Y.}, \au{Fantuzzi, G.}, \au{Papachristodoulou, A.},
  \au{Goulart, P.~J.} \& \au{Wynn, A.}} \yr{2017{\natexlab{{\em b\/}}}} {Fast
  ADMM for semidefinite programs with chordal sparsity}.  \bt{In {\em Proc.
  2017 Am. Control Conf.\/}},  \pg{pp. 3335--3340}. Seattle, USA.

\bibitem[Zuiderweg \& Harmens(1958)]{Zuiderweg1958}
{\sc \au{Zuiderweg, F.~J.} \& \au{Harmens, A.}} \yr{1958}  \at{{The influence
  of surface phenomena on the performance of distillation columns}}.  \jt{Chem.
  Eng. Sci.}  \bvol{9}~(2-3),  \pg{89--103}.

\end{thebibliography}

\end{document}